\documentclass[10pt,journal,compsoc]{IEEEtran}
\usepackage[cmex10]{amsmath}
\usepackage{graphicx}
\usepackage{epstopdf}
\usepackage{listings}
\usepackage{color}

\usepackage{array}
\usepackage{booktabs}
\usepackage{multirow}

\usepackage{amssymb}
\usepackage{gensymb}
\usepackage{color}
\usepackage{bm}

\usepackage{threeparttable} 
\usepackage{mathtools}

\DeclareMathOperator*{\argmin}{\arg\!\min}

\usepackage[utf8]{inputenc}
\usepackage[english]{babel}
 
\usepackage{amsthm}
 
\theoremstyle{definition}
\newtheorem{definition}{Definition}
 
\theoremstyle{remark}

\definecolor{myblue}{rgb}{0,0,0.9}
\definecolor{mygray}{rgb}{0.9,0.9,0.9}
\definecolor{mymauve}{rgb}{0.58,0,0.82}
  
\ifCLASSINFOpdf
\else
\fi

\begin{document}

\title{Modeling Attack Resilient Reconfigurable Latent Obfuscation Technique for PUF based Lightweight Authentication}


%
%
%
%
%
%
%
%

\author{Yansong~Gao,~\IEEEmembership{Student Member,~IEEE,}
~Said F.~Al-Sarawi,~\IEEEmembership{Member,~IEEE,}
~Derek~Abbott,~\IEEEmembership{Fellow,~IEEE}
~\\Ahmad-Reza Sadeghi,~\IEEEmembership{Member,~IEEE,} and Damith C.~Ranasinghe,~\IEEEmembership{Member,~IEEE,}
        
\IEEEcompsocitemizethanks{\IEEEcompsocthanksitem Y.~Gao, S.~F.~Al-Sarawi, D.~Abbott are with School of Electrical and Electronic Engineering, The University of Adelaide, SA 5005, Australia
\{yansong.gao, said.alsarawi, derek.abbott\}@adelaide.edu.au}
\IEEEcompsocitemizethanks{\IEEEcompsocthanksitem AR.~Sadeghi is with System Security Lab, Technische Universitat Darmstadt, Darmstadt 64289, Germany. ahmad.sadeghi@trust.tu-darmstadt.de}
\IEEEcompsocitemizethanks{\IEEEcompsocthanksitem Y. Gao and D.~C.~Ranasinghe are with Auto-ID Labs, School of Computer Science, The University of Adelaide, SA 5005, Australia. damith.ranasinghe@adelaide.edu.au}}
 
\IEEEtitleabstractindextext{
\begin{abstract}
Physical unclonable functions (PUFs), as hardware security primitives, exploit manufacturing randomness to extract hardware instance-specific secrets. One of most popular structures is time-delay based Arbiter PUF attributing to large number of challenge response pairs (CRPs) yielded and its compact realization. However, modeling building attacks threaten most variants of APUFs that are usually employed for strong PUF-oriented application---lightweight authentication---without reliance on the securely stored digital secrets based standard cryptographic protocols. In this paper, we investigate a reconfigurable latent obfuscation technique endowed PUF construction, coined as OB-PUF, to maintain the security of elementary PUF CRPs enabled authentication where a CRP is never used more than once. The obfuscation---determined by said random patterns---conceals and distorts the relationship between challenge-response pairs capable of thwarting a model building adversary needing to know the exact relationship between challenges and responses. A bit further, the obfuscation is hidden and reconfigured on demand, in other words, the patterns are not only invisible but also act as one-time pads that are only employed once per authentication around and then discarded. As a consequence, the OB-PUF demonstrates significant resistance to the recent revealed powerful Evaluation Strategy (ES) based modeling attacks where the direct relationship between challenge and response is even not a must. The OB-PUF's uniqueness and reliability metrics are also systematically studied followed by formal authentication capability evaluations.
\end{abstract}

}

\maketitle
\IEEEpeerreviewmaketitle
\section{Introduction}\label{Introduction}
\IEEEPARstart{M}{ost} current security solutions resort to storing a digital secret key in non-volatile memory (NVM) that remains hidden to an adversary. However, this requirement is difficult to realize in practice. Firstly, invasive
techniques, e.g., semi-invasive methods and side channel attacks, may extract valuable key
information \cite{gassend2002silicon,anderson2008security}. Secondly, malware such as Trojan horses or viruses can read
out and transfer keys unbeknownst to users~\cite{anderson2008security}. Furthermore, lightweight and mobile devices such as smart cards, Internet-of-Things (IoT) devices and radio-frequency identification (RFID) labels may have no dedicated room for key protection as functionality and cost consideration mostly dominate security requirements in commercial scenarios~\cite{anderson2008security}. 

The PUF offers a cost-effective alternative to building security services such as authentication or identification without the need to store secure keys in NVM~\cite{suh2007physical,katzenbeisser2012pufs,gao2016obfuscated}. A PUF extracts secrets on demand from unclonable and uncontrollable process variations resulting from manufacturing processes when the PUF is stimulated by an input signal (challenge).

Among all silicon based PUFs, Arbiter PUF (APUF) \cite{suh2007physical,lim2004extracting} receives extensive attentions as it holds the promise to serve as a strong PUF that yields an exponential number of challenge response pairs (CRPs). This desirable property enables a lightweight authentication mechanism based on CRPs. Though recent modeling attacks threaten such an elementary strong PUF-oriented application~~\cite{ruhrmair2010modeling,ruhrmair2013puf,becker2015gap,becker2015pitfalls}, the pursuit of a strong PUF-oriented lightweight entity authentication circuit has persisted and such a realization is regarded as an open problem in the face of plausible modeling attacks, \cite{roel2012physically,becker2015pitfalls,yulockdown,delvaux2015survey}. The most recent work from Yu {\it et al.} use a lockdown technique to prevent modeling attacks by enforcing and upper-bounding the available number of CRPs by an adversary in a trade-off of the available authentication rounds~\cite{yulockdown}.

We continue the investigation of a such strong PUF to support a lightweight authentication protocol and aim to eliminate the need to limit authentication rounds in~\cite{yulockdown} while showing heuristic security against modeling attacks. This article builds upon our preliminary work~\cite{gao2016obfuscated} to construct an OB-PUF. The OB-PUF obscures the relationship between the challenges and responses using a selected pattern vectors---shall be clear in Section ~\ref{ChallengePreProAndRec}---securely generated by the trusted party---server---and the OB-PUF embedded device. By exploiting the proposed reconfigurable obfuscation technique, the reconfigurable OB-PUF shows significantly increased resilience to modeling attacks, even the most powerful ES attacks where knowledge of direct relationship between challenge and response is not a must. We summarize the contributions of this work below:
\begin{itemize}
 \item We hide and reconfigure the obfuscation---determined by the pattern vectors---between the challenge and response, this prevents recent revealed powerful ES based modeling attacks. 
 \item We further implement response obfuscation besides the challenge obfuscation already considered in our previous work~\cite{gao2016obfuscated} in order to balance the Hamming weight of the obfuscated responses, which eliminates potential ES attacks based on Hamming weight enabled fittest objective as detailed in Section~\ref{Sec:HWESAttack}.  
 \item We develop a server-aided authentication mechanism that exploits the resourcefulness of a server to not only conduct successful authentication but reconfigure the obfuscation between the challenge and response on demand. We evaluate the authentication capability of OB-PUF primitives through a systematic study of its inter-distance and intra-distance, and subsequently, its false acceptance rate and false rejection rate in the authentication mechanism. In comparison with the most recent work~\cite{yulockdown}, the OB-PUF unlimits the available authentication rounds. 
  
\item We evaluate the heuristic security of reconfigurable OB-PUF primitives against modeling attacks, specifically, recently revealed powerful ES attacks. 
\end{itemize}
The remainder of the paper is organized as follows: Section~\ref{RelWork} introduces background and related work. Section~\ref{ChallengePreProAndRec} presents the proposed challenge-response obfuscation method and the corresponding method to recover the challenge and response and details their implementations. Section~\ref{Sec:ApplicationAuthentication} systematically analyses the authentication power/capability of the OB-PUF under different cost, security and performance settings; Section~\ref{Analysis} executes modeling attack tests to evaluate the enhanced security of reconfigurable OB-PUFs through extensive studies; and Section~\ref{Conclusion} concludes this paper.

\section{Background and Related Work}\label{RelWork}

\subsection{Background}

\subsubsection{\bf Physical Unclonable Functions}
When a PUF is stimulated by a challenge (input), $\bf C$, a corresponding response (output), $\bf R$, will be generated and determined by $f({\bf C})$, where $f(\cdot)$ is a physical function that is unique and inherent to each device. Given the same challenge, $\bf C$, different PUFs with the same design produce diverse responses, $\bf R$. The challenge, $\bf C$, and its corresponding response, $\bf R$, are commonly referred to as a challenge response pair (CRP). Over the years, a number of PUF structures have been proposed, built and analyzed. Popular PUF designs include \emph{time delay based} Arbiter PUF (APUF)~\cite{gassend2004identification,lee2004technique,lin2012design}, Ring-Oscillator PUF \cite{suh2007physical} (RO-PUF); \emph{Memory-based} PUFs leveraging device mismatch such as SRAM PUF~\cite{holcomb2009power,zeitouniremanence,saleem2016run}. Comprehensive reviews of different PUF architectures are referred to~\cite{herder2014physical,roel2012physically,gao2015emerging}. 

Examples of PUFs that build upon linear additive units include APUF~\cite{gassend2002silicon,suh2007physical,lim2005extracting} and $k$-sum (RO-sum) PUF~\cite{yu2011lightweight,herder2014physical}. They have one key desirable feature, generating exponential number of CPRs. Further, the hardware implementation is simple and the area overhead is small, especially the APUF. However, the major shortcoming is their vulnerability to modeling attacks. From a modeling attack perspective, both architectures have the same topology (constructed from linear additive blocks), therefore in this paper, the APUF is used for demonstration and description. Our work exploits the vulnerability of APUFs to modeling attacks. Therefore, we introduce the APUF next.
\subsection{Modeling APUF}
\begin{figure} [t]
\centering
\includegraphics[trim=0 0 0 0,clip,width=0.4\textwidth]{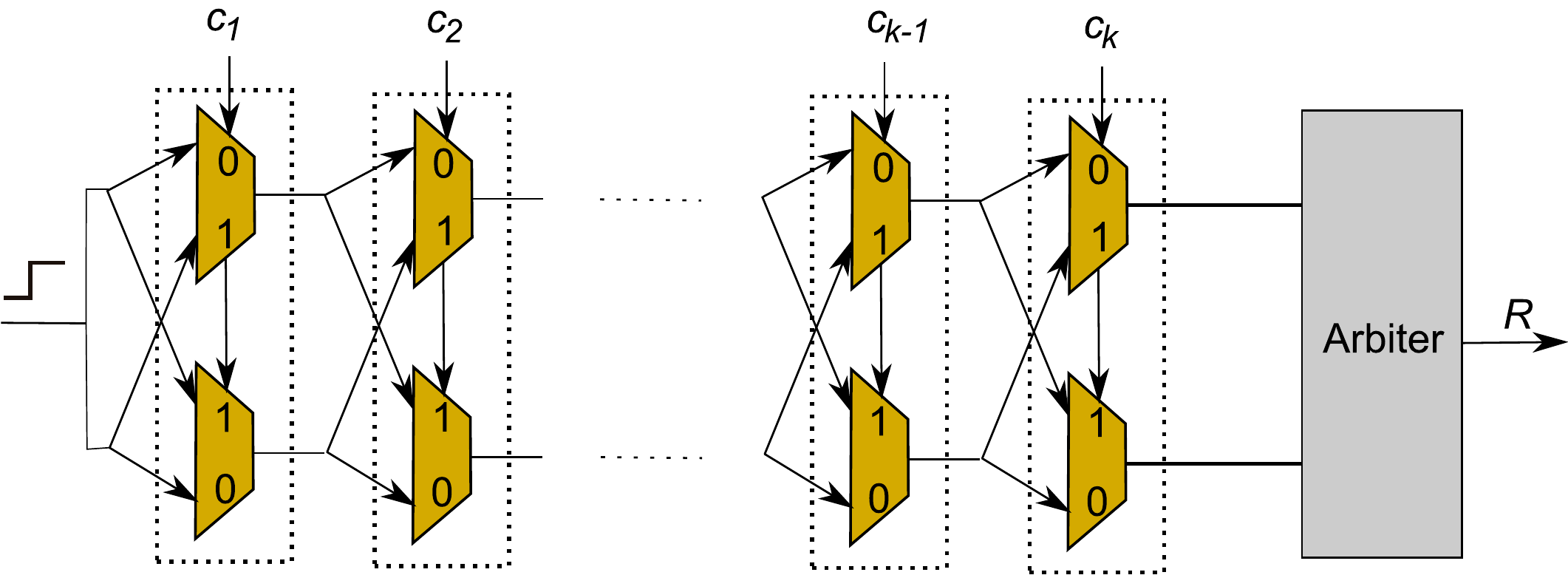}
\caption{An arbiter PUF (APUF) circuit.}
\label{APUF}
\end{figure}

The APUF consists of $k$ stages of two 2-input multiplexers as shown in Fig.~\ref{APUF}, or any other units forming two signal paths. To generate a response bit, a signal is applied to the first stage input, while the challenge $ \bf C $ determines the signal path to the next stage. The input signal will race through each multiplexer path (top and bottom paths) in parallel with each other. At the end of the APUF architecture, an arbiter, e.g., a latch, determines whether the top or bottom signal arrives first and hence results in a logic `0' or `1' accordingly.

It has been shown that an APUF can be modelled via a linear additive model because a response bit is generated by comparing the summation of each time delay segment in each stage (two 2-input multiplexers) depending on the challenge $\bf C$, where $\bf C$ is made up of ($ c_1|| c_2||  ...|| c_k $)~\cite{ruhrmair2013puf,lim2004extracting,ruhrmair2010modeling}. Following the notations in~\cite{ruhrmair2013puf,ruhrmair2010modeling}, the final delay difference $ t_{\rm dif} $ between these two paths is expressed as:
\begin{equation}\label{Eq:t_dif}
 t_{\rm dif} = {{\boldsymbol \omega}}^T {\bf {\Phi}} ,
\end{equation}

where $ {\boldsymbol \omega} $ and $\bf{ {\Phi}} $ are the delay determined vector and the parity vector, respectively, of dimension $k+1$  as a function of $\bf C$. We denote ${\sigma}_i^{1/0} $ as the delay in stage $i$ for the crossed ($ c_i=1 $) and uncrossed ($ c_i=0$) signal path through the multiplexers, respectively. Hence ${\sigma}_i^{1} $ is the delay of stage $i$ when $ c_i=1 $, while ${\sigma}_i^{0}$ is the delay of stage $i$ when $ c_i=0 $. Then
\begin{equation}
{{{\boldsymbol \omega}}}=({\omega}^1, {\omega}^2~...~ {\omega}^k, {\omega}^{k+1})^T,
\end{equation}
where ${\omega}^1= {{\sigma}_1^{0}-{\sigma}_1^{1} \over 2}$, $ {\omega}^i = {{\sigma}_{i-1}^{0} + {\sigma}_{i-1}^{1} + {\sigma}_{i}^{0} - {\sigma}_{i}^{1} \over 2} $ for all $ i=2,...,k $ and $ {\omega}^{k+1}={ {\sigma}_{k}^{0} + {\sigma}_{k}^{1} \over 2}$, also  
\begin{equation}\label{Eq:ChallengeFeature}
\bf{{\Phi}} (\bf{{C}}) = ({\Phi}^1(\bf{{C}}),...,{\Phi}^k(\bf{{C}}),1)^T,
\end{equation}
where $ {\bf {\Phi}^j}(\bf{{C}})={ {\rm \Pi}}_{\it i=j}^{\it k}({\rm 1-2}{\it c}_i) $ for $ j=1,...,k $. This $ \bf{{\Phi}} (\bf{{C}}) $ is also referred as the {\it challenge feature}.

Here we can see that the ${\boldsymbol \omega} $ encodes the delays for the subcomponents in the APUF stages, while the $ \bf{{\Phi}} $ is a challenge feature as a function of $ c_1,...,c_k $. The delay difference, $ t_{\rm dif} $, is the inner product of $ {\boldsymbol \omega} $ and $ \bf{{\Phi}} $. If $ t_{\rm dif}$ is greater than 0, the response bit is `1', otherwise, the response bit is `0'. 
\subsection{CMA-ES Attacks}\label{Sec:CMA-ESAttack}
Recently, Becker {\it et al.} extensively investigated the CMA-ES based modeling attacks \cite{becker2015gap,becker2015pitfalls}. In general, to attack an APUF, initially a randomly generated  ${\boldsymbol \omega} $, that is a $ k+1 $ length delay time vector, imitates the time delay model of an APUF. Such a ${\boldsymbol \omega} $ serves as a parent, many offspring delay vectors are generated based on the parent and some random mutations, e.g., random modifications of the parent delay vector. According to a fitness metric, the best child which maximize the fitness metric is determined and served as the parent for the next generation. This process is repeated untill a delay time vector of sufficient accuracy is formed. 

In general, a reasonable or good fitness metric must be defined---evaluating the goodness of the solution---to guide the ES algorithm to the best delay time vector despite the absence of a direct relationship between challenges and responses. 
\subsection{Related Work}
There are several approaches to increase the complexity of the task faced by an adversary to perform modeling attacks. One approach is to add nonlinearity, such as XOR-APUFs~\cite{lim2004extracting,ruhrmair2010modeling} or usage of Feed Forward Arbiter PUF~\cite{lim2004extracting,majzoobi2008testing} and the Lightweight Secure PUF~\cite{maj2008lightweight}. Unfortunately, increasing the complexity of model building attacks through the integration of more non-linear elements also significantly decreases the reliability of the PUF~\cite{lim2004extracting,ruhrmair2010modeling,ruhrmair2013puf}. Urich {\it et al.} show that all above strong PUF structures can be successfully attacked up to a certain size and complexity when knowing the exact relationship between challenges and responses~\cite{ruhrmair2010modeling,ruhrmair2013puf}.

Unlike the previous solutions, two recent proposals~\cite{rostami2014robust,yu2014noise} have demonstrated an alternative approach by hiding the direct relationship between a given challenge $\bf C$ and its response $\bf R$.  These two proposals in fact make use of the modeling attacks to build accurate models of APUFs by the trusted authority---the server---and then destroy the direct access to query direct challenge and response pairs, where only distorted challenge and response pairs are exposed afterwards~\cite{rostami2014robust,yu2014noise,majzoobi2012slender}.

In the first proposal~\cite{rostami2014robust,majzoobi2012slender}, the PUF integrated device, coined as Slender PUF, only reveals a subset of responses generated on the device for authentication, while the server can search and match the received substring since it can emulate PUF responses with its PUF model. In detail, a subset string of a full-length PUF response is randomly selected by the Slender and is subsequently padded with a randomly generated string to ensure the same response bit length expected from the PUF. The Slender PUF then sends the post-processed response to the server that uses a specific recovery method to discover the randomly selected subset string and decides the authenticity of the Slender PUF held by the end-user. In the second proposal~\cite{yu2014noise}, a decimation technique is developed to randomly elminiate response bits at the end-user with a PUF integrated device (prover) and a corresponding response recovery method is developed at the server to establish the validity of the response bit string. 

In summary, these two approaches efficiently conceal the challenge and response relationship to prevent an adversary from building an accurate model of a PUF by observing those visible CRPs. Although such an obfuscation mechanism in those two works efficiently prevent conventional modeling attacks usually requiring direct relationship between a challenge and a response~\cite{ruhrmair2010modeling,ruhrmair2013puf}. However, Becker {\it et al.} recently demonstrated that they are vulnerable to Evaluation Strategy (ES) based methods, in particular covariance matrix adaptation ES (CMA-ES) as detailed in Section~\ref{Sec:CMA-ESAttack}.

We can also consider the controlled PUF~\cite{gassend2002controlled} employing an obfuscation technique. It, however, requires relatively expensive on-chip error correction, the storage of associated helper data and subsequently hash logic for obfuscation, which are not needed in~\cite{yu2014noise,rostami2014robust,majzoobi2012slender}. In addition, the employment of helper data actually places the controlled PUF under potential security threats from ES attacks~\cite{becker2015pitfalls}.

A recent work from Yu {\it et al.} restricts the available CRPs to an adversary using a lockdown technique~\cite{yulockdown}. In this context, an adversary is faced with the diffculty of gaining enough training materials---CRPs---to learn an accurate model. This still allows lightweight authentication applications by upper-bounding available number of secure authentication rounds over the device's lifetime. Notably, this lockdown technique also takes advantage of the resource-rich server to build underlying APUF models during the secure enrollment phase, similar to ~\cite{rostami2014robust,yu2014noise,majzoobi2012slender}. Also noting that all works~\cite{rostami2014robust,yu2014noise,majzoobi2012slender,yulockdown} reap benefits from exploiting on-chip nonce or random number generator (RNG). 

We build upon the desirable features of previous works~\cite{yu2014noise,rostami2014robust,majzoobi2012slender,yulockdown} such as: i) enrolling the APUF model during the secure enrollment phase rather characterizing CRPs and storing them, especially when large number of authentication rounds are required; ii) obfuscation to remove direct relationship between CRPs; iii) device nonce. Our obfuscation method distorts relationship between a challenge and a response using a latent pattern that can be reconfigured (nonce). Such a technique results in an OB-PUF design that: i) shows heuristic security against most powerful ES attacks \cite{becker2015gap,becker2015pitfalls} ; ii) eschews the need to restrict number of authentication rounds in~\cite{yulockdown}; iii) removes reliance on on-chip error-correction logic, computing and storing associated helper data and subsequent hashing logic as in~\cite{gassend2002controlled}. 
\section{Challenge-Response Obfuscation and Recovery} \label{ChallengePreProAndRec}


\begin{figure}
	\centering
	\includegraphics[trim=0 0 0 0,clip,width=0.40\textwidth]{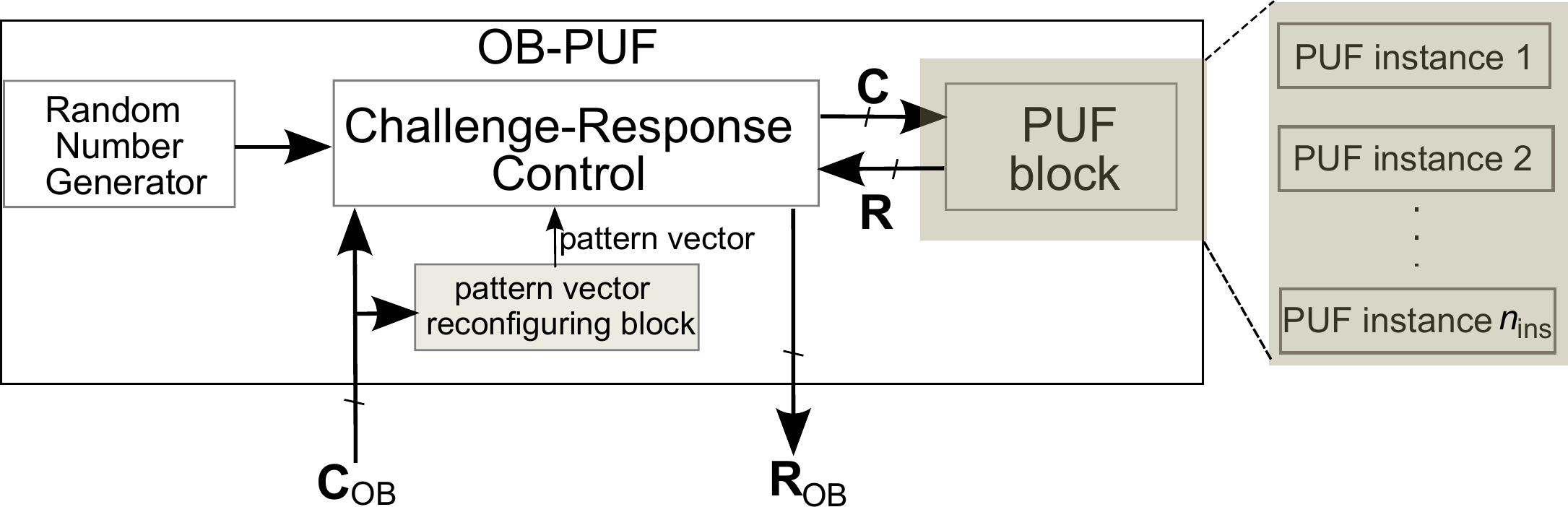}
	\caption{The OB-PUF structure.}
	\label{ChallengePreProcessing}
\end{figure}
Fig.~\ref{ChallengePreProcessing} illustrates the overall architecture of our challenge-response obfuscated PUF---the OB-PUF. The OB-PUF is built upon the PUF block consisting of $ n_{\rm ins} $ APUFs that share the same full length challenge $ \bf C $. The challenge stimulated to the OB-PUF is a {\it partial challenge} $ \bf C_{\rm OB} $ of $ k-m $ bits and the generated {\it obfuscated response} $ {\bf R}_{\rm OB} $ is of $ n_{\rm ins} $ bits. A partial challenge is directly applied to the {\it challenge-response control} block within the OB-PUF. The random number generator (RNG) is used to select a {\it pattern vector}. A pattern vector refers to the values and positions of $ m $ bits used to insert into a partial challenge to form a full length challenge, and the string that is XOR-ed with an $ n_{\rm ins} $ bits response $\bf R $ from the PUF block to realize the obfuscated response. 

We first use a fixed pattern vector to make the descriptions more readable without affecting analyses on the OB-PUF except modeling attacks analyses. The elaborate way in which the pattern vector, more precisely, the inserted values investigated in this paper, are hidden and reconfigured on demand by using APUF responses produced within the pattern vector reconfiguring block will be fully described in Section~\ref{Sec:ReconfigurePattern}.

\begin{figure}
	\centering
	\includegraphics[trim=0 0 0 0,clip,width=0.4\textwidth]{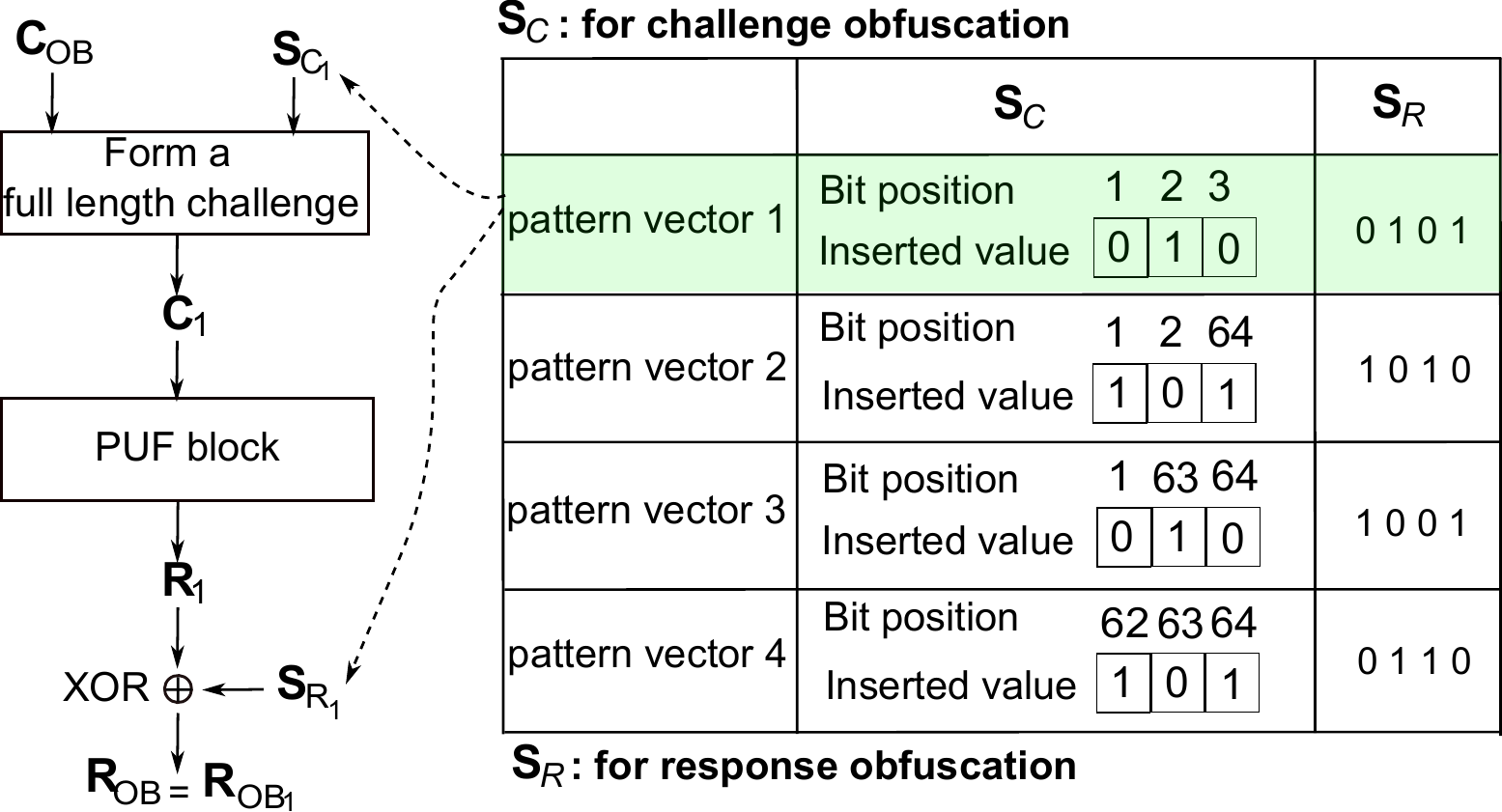}
	\caption{An example of challenge-response obfuscation. The {\it pattern vector} $ ({\bf S}_{\rm C_{1}},{\bf S}_{\rm R_{1}})$ is randomly chosen by the RNG. The $ {\bf S}_{\rm C_{1}} $ vector determines the formation of the full length challenge $ {\bf C}_1 $ by controlling positions and values of the $ m=3 $ bits that can be inserted into a partial challenge $ \bf C_{\rm OB} $. The $ {\bf S}_{\rm R_{1}}$ vector with length of $ n_{\rm ins} $ is XOR-ed with $ {\bf R}_1 $. In general, it can be seen that one $ {\bf C}_{\rm OB} $ can result in one of $ p=4 $ possible full length challenges, $ {\bf C}_1,...,{\bf C}_p $, and hence one of $ p $ possible corresponding responses, $ {\bf R}_1,...,{\bf R}_p $. The response ${\bf R}_i$ is further obfuscated by XOR-ing with a specific $ {\bf S}_{\rm R_{i}}$ vector to obtain the obfuscated response $ {\bf R}_{\rm OB_i}$, according to the selected pattern vector $ i $. The observable obfuscated response $ {\bf R}_{\rm OB}={\bf R}_{\rm OB_i} $, determined by the chosen pattern vector  $ ({\bf S}_{\rm C_{i}},{\bf S}_{\rm R_{i}})$, is sent from the OB-PUF. Note pattern vectors are latent and reconfigurable, specifically, this work focuses on reconfiguring the inserted values in $ {\bf S}_{C} $ vector. 
	}
	\label{InsertBitValues}
\end{figure}
As an example, consider an OB-PUF with four pattern vectors shown in Fig.~\ref{InsertBitValues}, where the PUF block consists of four APUFs. A partial challenge is applied to the {\it challenge-response control block} of the OB-PUF. Assuming the RNG selects pattern vector 1, the full length challenge $ {\bf C}_1 $ is formed by inserting logic values of \{`0', `1', `0'\} into the partial challenge at the respective bit positions of 1, 2, and 3. Subsequently, the formed full length challenge $ {\bf C}_1 $ from the challenge-response control block is applied to the PUF block. Recall, the PUF block comprises of four basic PUF instances implemented in parallel. These four basic PUF instances share the same full length challenge $ {\bf C}_1 $ in order to generate a 4-bit response simultaneously---the reason is described in Section~\ref{Sec:DistDec}. After the response ${\bf R}_1 $ is generated from the PUF block, it is further processed by the challenge-response control block. Here ${\bf R}_1 $ is XOR-ed with the string `0101' in order to obtain the response $ {\bf R}_{\rm OB_1} $. Then $ {\bf R}_{\rm OB_1} $ is sent out from the challenge-response control block as the obfuscated response $ {\bf R}_{\rm OB} $ of the OB-PUF. 

The central concept behind the challenge-response obfuscation is the hiding of the pattern vector chosen by the RNG. As a consequence, the full length challenge ${\bf C}_i, i\in\{1,...,p\},$ and the corresponding response $ {\bf R}_i $ are not only hidden from the adversary but also the server, but only the server can recover it by implementing the challenge-response recovery method, which is described in detail in Section \ref{Sec:ChallengeRecMethod}.

In the following subsections, first, we introduce relevant notations and definitions. Secondly, we introduce the {\it challenge-response obfuscation} employed to obfuscate the full length challenge applied to the PUF block inside the OB-PUF and the response from the underlying PUF block. Thirdly, we present the {\it challenge-response recovery} method to enable the server to discover the selected pattern vector by observing the obfuscated response, and hence, the obfuscated full length challenge and its corresponding response. 
\subsection{Preliminaries}
Before delving into the details and evaluation of the OB-PUF, we give a number of useful definitions and then introduce several notations where the parameters used in the following descriptions are listed in Table~\ref{tab:para}.
\begin{table} 
	\centering 
	\caption{Description of parameters}
	\resizebox{0.40\textwidth}{!}{
	\begin{tabular}{c |c} 
		\toprule 
		Parameter & Description\\ \hline 
		\midrule 
		$ k $ & \begin{tabular}{@{}c@{}} Length of a PUF challenge vector (stages of APUF)\end{tabular}  \\\hline 
		$ n_{\rm ins} $ & \begin{tabular}{@{}c@{}} Number of PUF instances implemented in \\ parallel on the PUF block of the OB-PUF\end{tabular}  \\\hline 
		$ m $& Number of eliminated challenge bits \\\hline 
		$ p $& Number of pattern vectors\\\hline
		$ \bf C_{\rm OB}$& \begin{tabular}{@{}c@{}} Partial challenge with length of $k-m$ \\bits applied to the OB-PUF\end{tabular}\\\hline
		$ \bf R_{\rm OB}$& Obfuscated response of $ n_{\rm ins} $-bit length from the OB-PUF\\\hline
		$ n $& \begin{tabular}{@{}c@{}} Number of challenge response pairs needed during\\ authentication to meet required false acceptance\\ rate and false rejection rate \end{tabular}\\\hline
		$ n_{\rm EER} $& \begin{tabular}{@{}c@{}} Threshold for obtaining equal false acceptance\\ rate and false rejection rate given a fixed $ n $\end{tabular} \\\hline
		$ n_{\rm mismatch} $&\begin{tabular}{@{}c@{}} A threshold that the server relies on to compare\\ the emulated response, $ {\bf R}_{\rm OB}^{\prime} $, from the server,\\ and the received response, $ {\bf R}_{\rm OB}$, from the \\physical OB-PUF, see {\it Definition \ref{def:nmismatch}} \end{tabular}\\ \hline
		
		$ {P}_{\rm pred} $& \begin{tabular}{@{}c@{}} The prediction accuracy of the PUF model\\ obtained by an adversary \end{tabular}\\\hline
		$ {P}_{\rm min} $& \begin{tabular}{@{}c@{}} The minimum prediction accuracy of thePUF model that \\ should be achieved by an adversary to break the PUF \end{tabular}\\\hline
		\bottomrule 
	\end{tabular}}
	\label{tab:para} 
\end{table}
As the Hamming distance (HD) and fractional Hamming distance (FHD) are often referred to in this work, we define them below, followed by definitions of the mean of pairwise HD and FHD.
\begin{definition}{\bf Hamming distance.}
For binary strings $ {\bf X}_1 $ and $ {\bf X}_2 $ with the same length $ l $, the HD between them is defined as:
\begin{equation}\label{eq:HD}
{f_{\rm HD}}({\bf X}_1, {\bf X}_2)=\sum_{i=1}^{l} {\bf X}_1\oplus {\bf X}_2.
\end{equation}
\end{definition}

\begin{definition}{\bf Fractional Hamming distance.}
Built upon (\ref{eq:HD}), the fractional Hamming distance (FHD) is defined as:
\begin{equation}\label{eq:FHD}
f_{\rm FHD}({\bf X}_1, {\bf X}_2)=\frac{f_{\rm HD}({\bf X}_1, {\bf X}_2)}{l}.
\end{equation}
\end{definition}

\begin{definition}{\bf Mean of pairwise HD/FHD.}\label{pairHD}
Given a collection $ {\bm {\mathcal X}} $ that consists of many binary strings, the mean of pairwise HD of $ {\bm {\mathcal X}} $ is defined as:
\begin{equation}
f_{\rm meanHD}(\bm {\mathcal X})={\rm mean}\{f_{\rm HD}({\bf X}_i,{\bf X}_j)\},
\end{equation}
where these two binary strings $ {\bf X}_i \in {\bm {\mathcal X}}, {\bf X}_j \in {\bm {\mathcal X}} $ and $ i\neq j $.

And the mean of pairwise FHD is defined as:
\begin{equation}
f_{\rm meanFHD}(\bm {\mathcal X})=\frac{f_{\rm meanHD}(\bm {\mathcal X})}{l}
\end{equation}
\end{definition}

Now consider the OB-PUF illustrated in Fig.~\ref{ChallengePreProcessing} and~\ref{InsertBitValues}, formally, we can define a pattern vector as a vector chosen from a set, $\{({\bm { S}_{\rm C_i}};{\bm { S}_{\rm R_i}})\} $, $ i\in \{1,...,p\} $. Here the $ {\bm {S}_{\rm C_i}} \in {\bm {\mathcal S}_{C}}$, where ${\bm {\mathcal S}_{C}} = \{{{\bm { S}_{\rm C_1}}},...,{\bm { S}_{\rm C_p}}\}$ determines the positions and values inserted into a partial challenge $ {\bf C }_{\rm OB} $ to form a possible full length challenge $ {\bf C}_i $. The $ {\bm { S}_{\rm R_i}} \in {\bm {\mathcal S}_{ R}}$, where ${\bm {\mathcal S}_{ R}} = \{{\bm { S}_{\rm R_1}},...,{\bm {S}_{\rm R_p}}\} $ and the length of $ {\bf S}_{\rm R_i} $ is $ n_{\rm ins} $, is the string that is XOR-ed with the response $ {\bf R}_i $ obtained from the PUF block in order to gain the obfuscated response.

There are $ p $ pattern vectors, therefore, there are $ p $ possible full length challenges given a partial challenge. Then given a pattern vector, $ i $, the formed full length challenge  $ {\bf C}_i $ applied to the PUF block is from a set $ {\bm {\mathcal C}} $, where $ {\bm {\mathcal C}} = \{{\bf C}_1,...,{\bf C}_p\} $. Consequently, the corresponding response $ {\bf R}_i \in {\bm {\mathcal R}}$, $ {\bm {\mathcal R}} = [{\bf R}_1,...,{\bf R}_p]$, is generated from the PUF block. The obfuscated response $ {\bf R}_{\rm OB_i} $ from the OB-PUF is a function of $ {\bf R}_i $, where $ {\bf R}_{\rm OB_i} = {\bf R}_i \oplus {\bm { S}_{\rm R_i}} $. Therefore, the final output $ {\bf R}_{\rm OB}= {\bf R}_{\rm OB_i}$ of the OB-PUF is from a collection of $ {\bm {\mathcal R}}_{\rm OB} $, where $ {\bm {\mathcal R}}_{\rm OB} = [{\bf R}_{\rm OB_1},...,{\bf R}_{\rm OB_p}] $. In other words, the final output---the observable obfuscated response---is actually determined by the randomly selected $ i_{\rm th} $ pattern vector.
\subsection{Challenge Obfuscation}\label{Sec:ChallengeOB}
The partial challenge intentionally eliminates $ m $ bits from a full length challenge. From a machine learning perspective, not only do we eliminate $ m $ features but also eliminate them from randomly selected positions. When such an incomplete input feature vector is employed to train a model, the missing features decrease the accuracy of the learned model. In Following, we describe the design of pattern vectors and realization of challenge obfuscation. 

As for the OB-PUF, a partial challenge gives $ p $ possible full length challenges, ${\bf C}_1,...,{\bf C}_p $. Hence, there are  $ p $ possible responses, $ {\bf R}_1,...,{\bf R}_p $. According to our experiments, if the mean of pairwise HD of possible full length challenges $ {\bm {\mathcal C}} $---see {\bf Definition \ref{pairHD}}---is small, the mean of pairwise HD of possible responses $ {\bm {\mathcal R}} $ will also be small. 

As an exemplary demonstration, we have randomly generated 5000 $ {\bf C}_{\rm OB} $ and form the four possible full length challenges according to one extreme case, where we inserted all values to the {\it first three positions} in each pattern vector---note that we do {\it not} insert the values to the positions that are optimal positions as shown in Fig.~\ref{InsertBitValues}. We got a 3.1\% mean of pairwise FHD of four possible full length challenges, consequently, the mean of pairwise FHD of four possible responses is only 1.1\%. In essence, recall (\ref{Eq:ChallengeFeature}), this extreme position selection remains $ ({\bf {\Phi}}^4(\bf{{C}}),...,{\bf {\Phi}}^{\it k}(\bf{{C}}),{\rm 1}) $ for all four possible full challenge fractures to be same. This creates a problem. Assume that the OB-PUF does not further implement response obfuscation and the mean of pairwise HD of $ {\bm {\mathcal C}} $ is very small. Given one possible response $ {\bf R}_i \in {\bm {\mathcal R}} $, for a $ {\bf C}_{\rm OB} $, is directly exposed as the response of OB-PUF. Now it does not matter which pattern vector is chosen to construct a full length challenge, an adversary can select any possible full length challenge from $ {\bm {\mathcal C}} $ and the exposed $ {\bf R}_i $ to train a model. This is because, by virtue of the small HD between full length challenges in $ {\bm {\mathcal C}} $, the mean of pairwise HD of possible responses $ {\bm {\mathcal R}} $ is also small. In other words, it is reasonable for an adversary to select any $ {\bf C}_j \in {\bm {\mathcal C}}, j\in \{1,...,p\},  $ to pair with the observable $ {\bf R}_i $ as a useful CRP to train a model.

To mitigate the above security concern, the mean of pairwise HD of possible responses $ {\bm {\mathcal R}} $ for a given partial challenge should be maximized. Alternatively, the mean of pairwise HD of possible full length challenges $ {\bm {\mathcal C}} $ should be maximized.
\begin{figure}
\centering
\includegraphics[trim=0 0 0 0,clip,width=0.47\textwidth]{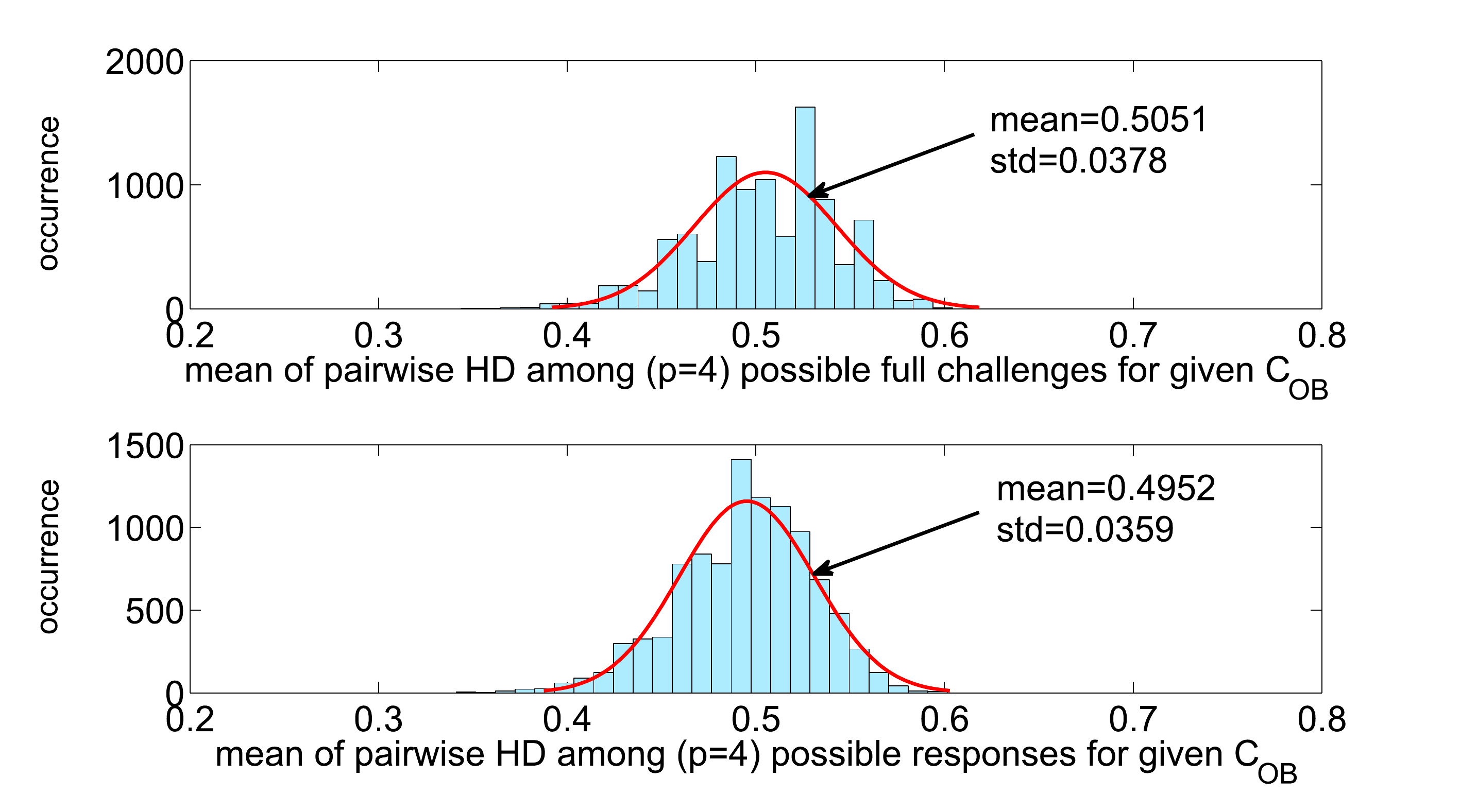}
\caption{The mean of pairwise FHD of $ {\bm {\mathcal C}} $ and $ {\bm {\mathcal R}} $ distribution.}
\label{MinimumHD}
\end{figure}

Then, it follows that to maximize the mean of pairwise HD of possible responses $ {\bm {\mathcal R}} $, the pattern vector used to generate full length challenges should be carefully designed. The four pattern vectors, $ p=4 $ as an example, listed in Fig.~\ref{InsertBitValues} is one strategy to maximize the mean of pairwise HD of possible full length challenges $ {\bm {\mathcal C}} $. 

\begin{figure}
\centering
\includegraphics[trim=0 0 0 0,clip,width=0.47\textwidth]{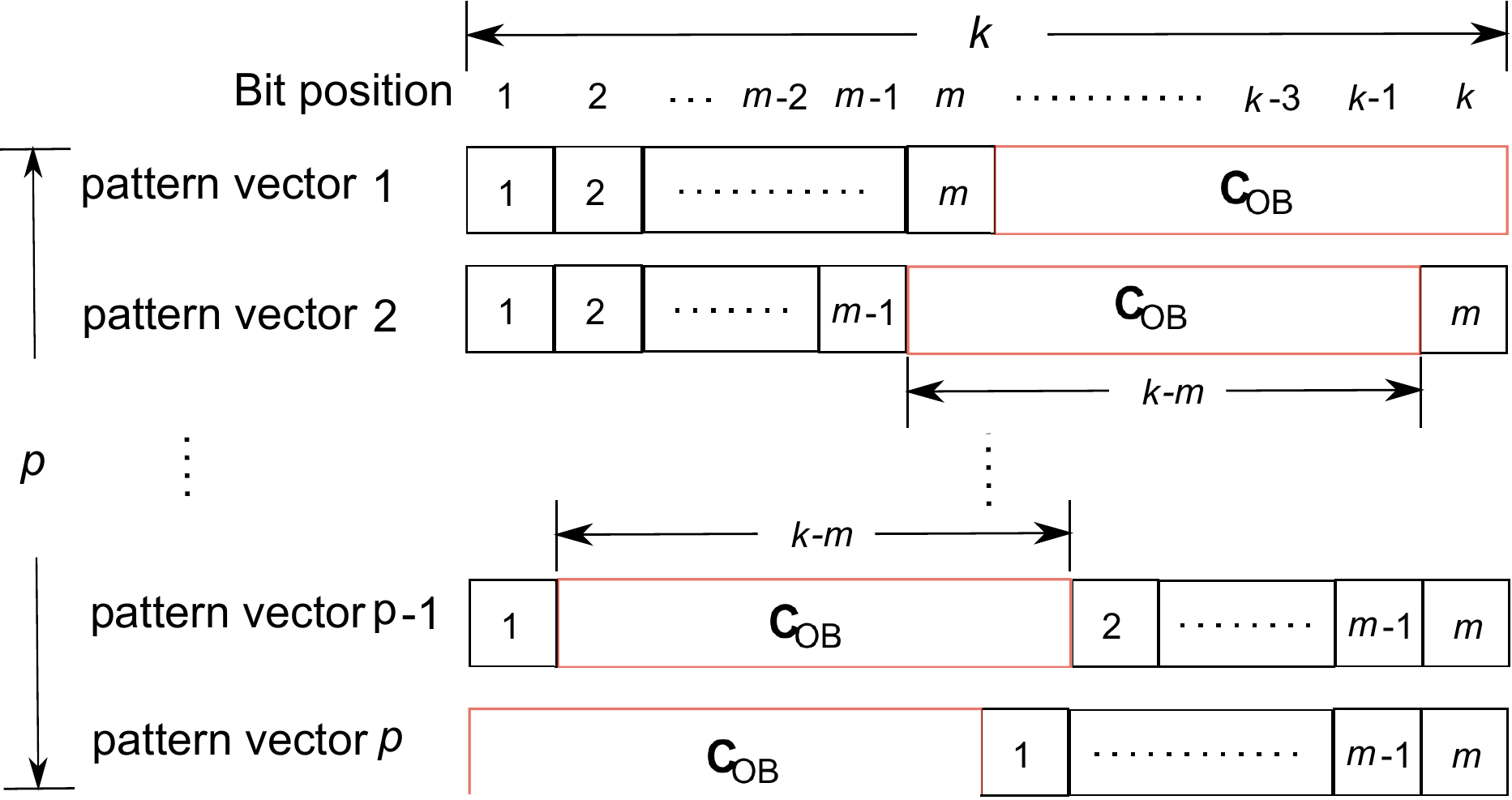}
\caption{Generalized pattern vector design method to maximize the mean of pairwise FHD of possible full length challenges $\bm {\mathcal C} $ given a randomly selected partial challenge $ \bf C_{\rm OB} $.}
\label{InsertBitGeneral}
\end{figure}

A generalized pattern vector design for challenge obfuscation is illustrated in Fig.~\ref{InsertBitGeneral}. In Fig.~\ref{MinimumHD}, it shows the distribution of the mean of pairwise FHD of possible full length challenges $ {\bm {\mathcal C}} $ and the distribution of the mean of pairwise FHD of corresponding responses $ {\bm {\mathcal R}} $. To generate the results in Fig.~\ref{MinimumHD}, we calculate the mean of pairwise FHD of possible full length challenge $ {\bm {\mathcal C}} $ and the corresponding possible responses $ {\bm {\mathcal R}} $ for each randomly generated partial challenge. The number of randomly generated partial challenges is 1,000,000, consequently, the distributions of the mean of pairwise FHD of possible full length challenges  $ {\bm {\mathcal C}} $ and the corresponding possible responses $ {\bm {\mathcal R}} $ for 1,000,000 partial challenges are acquired, where $ k=64 $ and $ n_{\rm ins}=64 $. It can be observed from Fig.~\ref{MinimumHD} that the mean of pairwise FHD of possible full length challenges $ {\bm {\mathcal C}} $ and the corresponding possible responses $ {\bm {\mathcal R}} $ are always in vicinity of 50\%, and, hence, indicating that the mean of pairwise FHD of $ {\bm {\mathcal C}} $ and $ {\bm {\mathcal R}} $ are maximized. Notably, a large mean pairwise FHD of $ \bm {\mathcal R} $---close to 50\%---helps increase the ability to discover the hidden full length challenge by a server or a verifier armed with a model of OB-PUF as will be discussed in Section~\ref{Sec:ChallengeRecMethod}.
\subsection{Response Obfuscation}\label{Sec:ResponseOB}
Response obfuscation further increases the mean pairwise HD of possible responses from an OB-PUF. Most importantly, it is employed to prevent an adversary with full access to the OB-PUF possibly discovering and employing those partial challenges, where the responses generated from an APUF given a partial challenge is only determined by delay difference of $ k-m $ stages and is agnostic to the selection of the other $ m $ stages.

As a motivating example, consider the first bit---the response generated by the first PUF instance---for all possible full length challenges, $ {\bf C}_1,...,{\bf C}_p $, given the same $ \bf C_{\rm OB} $, there is a probability of $ \frac{1}{2^p} $ that the first PUF instance produces the same response (`0'/`1') for all $ p $ possible pattern vectors. Here, if an adversary has full access to the OB-PUF, then the adversary can apply the same $ \bf C_{\rm OB} $ to the same OB-PUF repeatedly, to observe the first bit, $ r_1 $, of response $ \bf R $ to discover partial challenges with a response that is agnostic to the pattern vector. In other words, the response of the first APUF is only determined by delay difference of $ k-m $ stages. If this occurs, then the first PUF instance keeps producing the same response (`0'/`1'), and the adversary can utilize any possible full length challenge  given such a partial challenge and the corresponding response (`0'/`1') to train a model. As a consequence, through multiple evaluations of the same partial challenge $ \bf C_{\rm OB} $ and utilizing those $ \bf C_{\rm OB} $ that always generate the same response, an adversary can sequentially model each APUF instance in the PUF-block.

Response obfuscation is implemented to address such a scenario. The response obfuscation control involves XOR-ing $ {\bf R}_i $ with a pre-designed string to generate $ {\bf R}_{\rm OB_i} $, where $ {\bf R}_{\rm OB_i}={{\bf R}_i}\oplus {\bf S}_{\rm R_i} $, as illustrated in Fig.~\ref{InsertBitValues}. The design of $ {\bm {\mathcal S}_{ R}} $ obeys two rules: i) maximizing the mean of pairwise HD of $ {\bm {\mathcal S}_{ R}} $ to be 50\%; and ii) ensuring the percentage of `1' is close to 50\% for any given bit position of $ {\bm {\mathcal S}_{ R}} $ corresponding to every pattern vector (see the example pattern vectors in Fig.~\ref{InsertBitValues}).

After response obfuscation is implemented, the adversary must guess the randomly selected pattern vector to derive the response $ {\bf R}_i $ to train a model even if the $ j_{\rm th}, j\in \{1,...,n_{\rm ins}\} $ bit in $ {\bf R}_{\rm OB_i} $ always presents the same value. The reason lies in the fact that the selected $ {\bf S}_{\rm R_i} $ needs to be found in order to derive the $ {\bf R}_i $. However, the selected $ {\bf S}_{\rm R_i} $ is invisible. Hence, those partial challenges that always generate the same obfuscated responses leak no information that can be exploited by an adversary to break the OB-PUF.

\subsection{Pattern Vector Design}
We can see from our discussions above that the security of challenge-response obfuscation relies on the design and random selection of pattern vectors. In summary:
\begin{enumerate}
\item The challenge obfuscation pattern vector $ {\bf S}_C $ must be designed to maximize the pairwise HD of possible full length challenges $ {\bf C}_1,...,{\bf C}_p $ for a given $ {\bf C}_{\rm OB} $. 
\item The response obfuscation pattern vector $ {\bf S}_R $ must be designed to maximize the mean of pairwise HD of $ {\bm {\mathcal S}_{ R}} $ to be 50\% and to ensure the percentage of `1' in a given bit position of $ {\bm {\mathcal S}_{ R}} $ corresponding to every pattern vector to be approximately 50\%.
\end{enumerate}

\subsection{Challenge-Response Recovery}\label{Sec:ChallengeRecMethod}

\begin{figure}[h]
 	\centering
 	\includegraphics[trim=0 0 0 0,clip,width=0.35\textwidth]{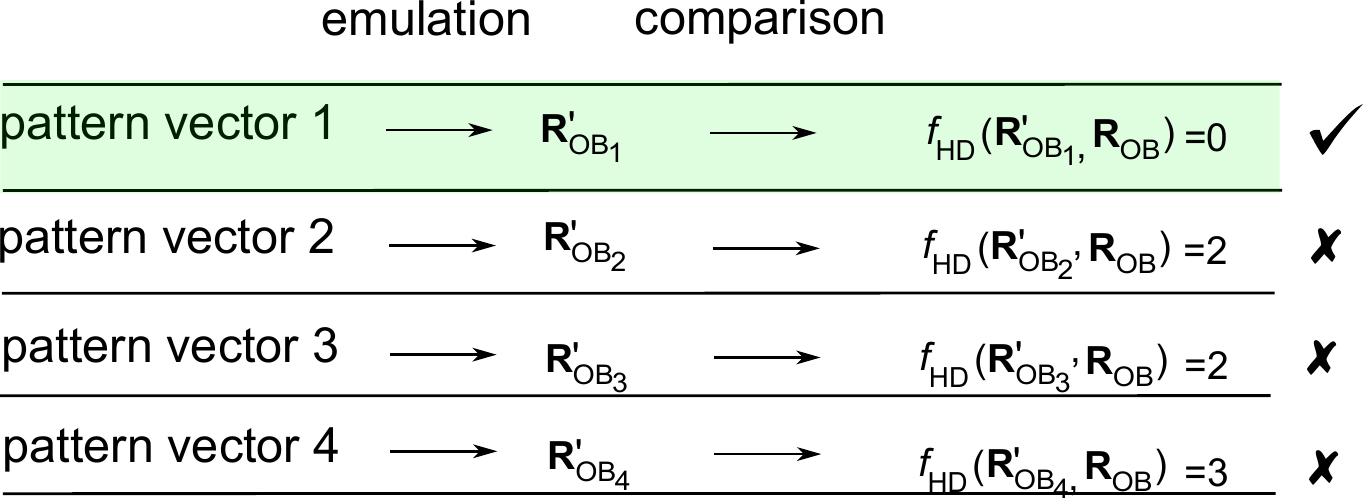}
 	\caption{Illustration of challenge-response recovery.}
 	\label{fig:PatternRecover}
 \end{figure}

When a server receives an obfuscated response, $ {\bf R}_{\rm OB} $, from an OB-PUF, the server can emulate all $ p $ possible obfuscated responses ${\bm {\mathcal R_{\rm OB}^{\prime}}}=[ {\bf R}_{\rm OB_1}^{\prime},...,{\bf R}_{\rm OB_p}^{\prime}]$, for a given $ {\bf C}_{\rm OB} $, according to the pattern vectors known by the server. 
As illustrated in Fig.~\ref{fig:PatternRecover}, the server compares each emulated obfuscated response $ {\bf R}_{\rm OB_i}^{\prime}, i\in\{1,...,p\} $ with the received $ {\bf R}_{\rm OB} $ from the user and accept if there exists one emulated obfuscated response $ {\bf R}_{\rm OB_i}^{\prime}$ that matches the received $ {\bf R}_{\rm OB}$. The matching criterion is detailed in Section~\ref{Sec:AuthenSecheme} along with the OB-PUF based authentication protocol.
 \section{Authentication Application and Analysis}\label{Sec:ApplicationAuthentication}
 In this section, first, we summarize a parameter-based authentication protocol applicable to the OB-PUF primitive. In addition, the authentication protocol can further harness a server-aided computation method to improve the OB-PUF's performance. Next, we introduce the general formula for establishing the authentication capability of PUFs and evaluate the crucial parameters, inter-distance and intra-distance, necessary to assess the authentication capability of our OB-PUFs. Then we quantitatively analyze the OB-PUF's authentication capability. At the end, we detail how to reconfigure the pattern vectors in authentication applications.
 
 \subsection{Server-Aided Parameter Based Authentication}\label{Sec:AuthenSecheme}
 
 {\bf Authentication Protocol.} A variant of parameter-based authentication protocol~\cite{yu2014noise} is described in detail below:
 \begin{description}
    \item[{\bf 1}:] A server gains a specific number of CRPs of each APUF in a PUF block to train a model and securely stores the parameter model associated with a given instance of an OB-PUF. This is usually referred to as the enrollment phase. Then the direct extraction of CRPs of APUF instances are destroyed, e.g., via fusing the extraction wires. 
    \item[{\bf 2}:] Whenever the prover requests an authentication from the server. The server randomly issues a ${\bf C}_{\rm OB} $ and sends it to the prover. The prover applies the ${\bf C}_{\rm OB} $ to the OB-PUF and obtains an $ {\bf R}_{\rm OB} $.
    \item[{\bf 3}:] After receiving $ {\bf R}_{\rm OB} $ from the prover, the server carries out the authentication by implementing the challenge-response recovery method. If one of the emulated obfuscated responses $ {\bf R}_{\rm OB}^{\prime} $ for a given $ {\bf C}_{\rm OB} $ matches the received obfuscated response $ {\bf R}_{\rm OB} $---see Section~\ref{Sec:ChallengeRecMethod}, the authentication is successful, and the authenticity of the physical entity is established; otherwise, if none of the emulated obfuscated responses $ {\bf R}_{\rm OB}^{\prime} $ matches to the received $ {\bf R}_{\rm OB} $, the authentication fails. A specific $ {\bf C}_{\rm OB} $ is used no more than once to prevent replay attacks.
 \end{description} 
 Notably, this parameter based authentication can further exploit resourcefulness of the server by implementing the following server-aided method. 
 
 {\bf Determining a Match.} In the example in Fig.~\ref{fig:PatternRecover}, a match is determined if the criterion $ f_{\rm HD}({\bf R}_{\rm OB}^{\prime},{\bf R}_{\rm OB})=0 $ is satisfied when performing the step 3 of the authentication protocol. This occurs when the pattern vector that the server selects for emulation is the same as the pattern vector randomly chosen in the physical OB-PUF. Otherwise, if none of the emulated $ {\bf R}_{\rm OB}^{\prime}$ matches the received $ {\bf R}_{\rm OB}$. This authentication around is rejected. In general, a match criterion does not always require $ f_{\rm HD}({\bf R}_{\rm OB}^{\prime},{\bf R}_{\rm OB})=0 $. As long as $ f_{\rm HD}({\bf R}_{\rm OB}^{\prime},{\bf R}_{\rm OB}) $ is lower than a mismatch tolerance $ n_{\rm mismatch} $ defined below, the server can accept the authenticity of the received obfuscated response.
  
\begin{definition}{\bf Mismatch tolerance $ n_{\rm mismatch} $.} \label{def:nmismatch}
 \\The $ n_{\rm mismatch} $ is the Hamming distance---number of mismatching bits---between an emulated obfuscated response $ {\bf R}_{\rm OB_i}^{\prime}$ and the received obfuscated response $ {\bf R}_{\rm OB}$ tolerated by a match criterion. A server accepts the received obfuscated response $ {\bf R}_{\rm OB}$ if $ f_{\rm HD}({\bf R}_{\rm OB}^{\prime},{\bf R}_{\rm OB}) \le n_{\rm mismatch}$.
  \end{definition}
  In other words, $ n_{\rm mismatch} $ describes tolerated number of flipped bits between the $ {\bf R}_{\rm OB_i}^{\prime}$ and $ {\bf R}_{\rm OB}$ when the server compares them. The $ {\bf R}_{\rm OB}$ is accepted  when the HD between one of $ p $ possible emulated $ {\bf R}_{\rm OB}^{\prime} $ and the received $ {\bf R}_{\rm OB} $ is no more than $ n_{\rm mismatch} $. We will show in Section~\ref{AuthenPower} that relaxing match criterion by selecting a large mismatch tolerance in fact expedites the authentication process.
  Clearly a single $ {\bf C}_{\rm OB} $ and $ {\bf R}_{\rm OB} $ pair, shortly an OB-CRP, is not adequate to authenticate a large population of OB-PUFs. One solution is to implement multiple OB-PUF instances sharing the same $ {\bf C}_{\rm OB} $ on a device, but this results into higher area and power overhead. The alternative approach of using multiple number of challenge response pairs is more practical, and is commonly employed by PUF based authentication protocols~\cite{lim2005extracting,roel2012physically}. In our authentication protocol, this implies that steps 2 and 3 are repeated multiple times. The goal with either approach is to generate adequate number of response bits to authenticate or identify a large number of OB-PUFs.
  \begin{figure}
  \centering
  \includegraphics[trim=0 0 0 0,clip,width=0.30\textwidth]{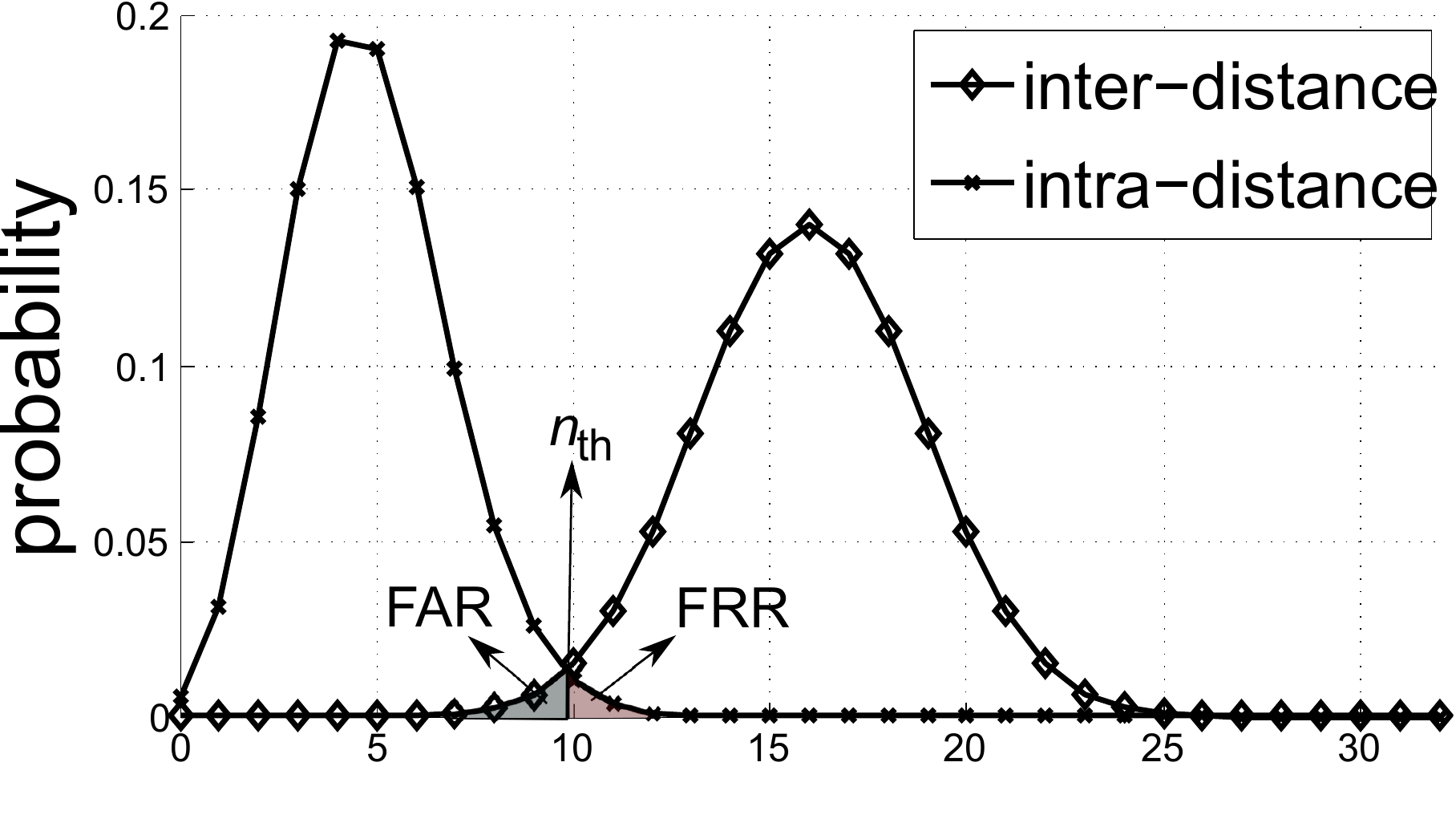}
  \caption{Distributions of intra-distance and inter-distance for 32-bit response.}
  \label{FARFRR}
  \end{figure}
  \subsection{Authentication Capability}
  Considering our need for multiple rounds or multiple OB-CRPs, the question that we need to address remains the number of OB-CRPs required to authenticate or identify an OB-PUF from a very large population of OB-PUFs.
  
  Starting with a conventional PUF, we can establish the number of CRPs needed to authenticate a PUF by considering the inter-distance and intra-distance used to describe the distribution statistics of PUF responses. We formally define the inter-distance and intra-distance following Roel's work~\cite{roel2012physically}.
  \begin{definition}{{\bf Inter-distance.}}\label{de:interDistance}
  The inter-distance is a random variable describing the distance between two PUF responses from different PUF instances subjected to the same challenge, hence,
  \begin{equation}\label{eq:interDistance}
  D_{\rm inter}={\rm dist}({\bf R}_A;{\bf R}_B)
  \end{equation}
  where $ {\bf R}_A$ and ${\bf R}_B $ are two responses from two randomly selected but distinct PUFs stimulated by the same challenge. The inter-distance measures the uniqueness of PUFs \cite{roel2012physically}.
  \end{definition}
  
  \begin{definition}{{\bf Intra-distance.}}\label{de:intraDistance}
  The intra-distance is a random variable describing the distance between two PUF responses from the same PUF instance and subjected the same challenge, hence,
  \begin{equation}\label{eq:intraDistance}
  D_{\rm intra}={\rm dist}({\bf R};{\bf R}^{\prime})
  \end{equation}
  where $ {\bf R}$ and ${\bf R}^{\prime} $ are two distinct responses from a randomly chosen PUF instance produced using the same challenge re-evaluated at two distinct times. The intra-distance describes a PUF's response reproducibility.
  \end{definition}
  
  The $ {\rm dist}(.;.) $ function above can be any well-defined and appropriate distance metric over the responses. In this paper, responses are always bit vectors and the used distance metric is Hamming distance or fractional Hamming distance---see Section \ref{Sec:ChallengeOB}. In Fig.~\ref{FARFRR}, we illustrate the distribution of the inter-distance and intra-distance of 32-bit length PUF responses. Considering that the evaluation of each CRP or each response bit is independent~\cite{lim2004extracting}, both inter- and intra-distance distributions are assumed to follow a binomial distribution, $ B(n,\hat{p}) $, where the binomial probability estimator of inter- and intra-distance distributions are ${\hat p}_{\rm inter} $ and $ {\hat p}_{\rm intra} $, respectively~\cite{lim2004extracting}. Then ${\hat p}_{\rm inter} $ is the probability that $ {\bf R}_A \ne {\bf R}_B $, and $ {\hat p}_{\rm intra} $ is the probability that $ {\bf R}\ne{\bf R}^{\prime} $~\cite{lim2004extracting}.
  
  We are particularly interested in the right tail of intra-distance distribution and the left tail of inter-distance distribution, because these two tails describe two undesirable errors in an authentication application: i) false rejection rate (FRR); and ii) false acceptance rate (FAR). 
  We can see that given an $ n $-bit response vector, FAR expresses the probability of misidentification and therefore a high FAR is a security weakness in the authentication mechanism. On the other hand, FRR expresses the probability of misrejection of an authentic PUF and therefore a high FRR will deem the authentication scheme impractical. Given both inter- and intra-distance follow binomial distributions, the FRR and FAR can be formally expressed in~\cite{lim2005extracting,roel2012physically}:
  \begin{eqnarray}
  \label{eq:FRR} {\rm FRR}& =& 1 - \sum\limits_{i=0}^{{n}_{\rm th}}{\binom{n}{i}}({ {{\hat p}_{\rm intra}}})^i{(1-{{ {\hat p}_{\rm intra}}})^{(n-i)}},\\
  \label{eq:FAR} {\rm FAR}& =& \sum\limits_{i=0}^{{n}_{\rm th}}{\binom{n}{i}}({ {\hat p}_{\rm inter}})^i{(1-
  {{\hat p}_{\rm inter}})^{(n-i)}}.
  \end{eqnarray}
  We can easily envision that a larger $ n_{\rm th} $ benefits FRR but increases the FAR, and vice versa. However, high values for both FRR and FAR are undesirable for authentication. Therefore, in practice, we want a balance between them. There exists a value for $ n_{\rm th} $ for which both FAR and FRR are equal. We refer to this value of $ n_{\rm th} $ as {\it equal error threshold}, $ n_{\rm EER} $, and when both error rates are equal, we refer to this as {\it equal error rate} (EER) following Roel's work in~\cite{roel2012physically}. For a discrete distribution, there may not be a value for $ {n}_{\rm EER} $ where FAR and FRR are exactly equal, and hence, $ {n}_{\rm EER} $ and EER are defined as in~\cite{roel2012physically}:
  \begin{eqnarray}
   {n}_{\rm EER} &=& \argmin_{{n}_{\rm th}} \{ {\rm max}\{{\rm FAR}({n}_{\rm th}),{\rm FRR}({n}_{\rm th}) \}\},\\
   {\rm EER} &=&  \max \{{\rm FAR}({n}_{\rm EER}),{\rm FRR}({n}_{\rm EER})\}.
  \end{eqnarray}
  Given a PUF class with binomial probability estimator ${\hat p}_{\rm inter} $ and $ {\hat p}_{\rm intra} $, we can find the minimum number of CRPs, $ n $, for ensuring an acceptable $ \rm EER $---typically in practice, a minimum requirement is a value lower than $ 10^{-6} $. During an OB-PUF authentication, the server sends out $ n $ partial challenges $ {\bf C}_{\rm OB} $ and compares the received $ n $ obfuscated responses $ {\bf R}_{\rm OB} $. Then, to evaluate an adequate value $ n $ to achieve an acceptable EER, we need to statistically evaluate the  $ {\hat p}_{\rm intra} $ and $ {\hat p}_{\rm inter} $ for our OB-PUF primitive.
  \subsection{The Binomial Probability Estimator $ {\hat p}_{\rm inter} $ of OB-PUF} \label{Sec:DistDec}
  Consider two 1-bit responses produced by applying the same challenge to two distinct conventional PUFs that are randomly chosen from a population. In general, the  $ {\hat p}_{\rm inter} $ of conventional PUFs is the probability that these two 1-bit responses are different. This is normally very close to 50\% since most PUFs can achieve a uniqueness very close to 50\% when the layout is carefully performed. Similarly, the $ {\hat p}_{\rm inter} $ of OB-PUF is the probability that two obfuscated responses are different, where these two obfuscated responses are generated by two randomly chosen OB-PUFs from a population given the same applied partial challenge.
  
  Formally, we can derive a generalized expression for $ {\hat p}_{\rm inter} $ of OB-PUFs. Consider that OB-PUF$_A  $ will generate a possible $ \bf R_{\rm OB_A} ~\in~[\bf R_{\rm OB_{A1}}, \bf R_{\rm OB_{A2}}, ..., \bf R_{\rm OB_{Ap}}]$, where $ p $ is the number of pattern vectors. Similarly, the OB-PUF$_B  $ will generate a possible $ \bf R_{\rm OB_B} \in [\bf R_{\rm OB_{B1}}, \bf R_{\rm OB_{B2}},..., \bf R_{\rm OB_{Bp}}]$. The $ {\hat p}_{\rm inter} $ of OB-PUFs is actually the probability $ \bf R_{\rm OB_A} \neq  \bf R_{\rm OB_B} $ given by:
  \begin{eqnarray}
  \nonumber&& P( {\bf R_{\rm OB_A}} \neq  {\bf R_{\rm OB_B}} )\\
  \nonumber&&= P(({\bf R_{\rm OB_{A1}}}\neq {\bf R_{\rm OB_B}})~\&~({\bf R_{\rm OB_{A2}}}\neq
  {\bf R_{\rm OB_B}}), \\
  \nonumber&&~~~...,~\&~({\bf R_{\rm OB_{Ap}}}\neq {\bf R_{\rm OB_B}}) )\\
  &&= P({\bf R_{\rm OB_{A1}}}\neq {\bf R_{\rm OB_B}})P({\bf R_{\rm OB_{A2}}}\neq {\bf R_{\rm OB_B}})\\
  \nonumber&&~~~...,P({\bf R_{\rm OB_{Ap}}}\neq {\bf R_{\rm OB_B}} ),
  \end{eqnarray}
  where
  \begin{eqnarray}
  \nonumber&& P( {\bf R_{\rm OB_{Ai}}} \neq  {\bf R_{\rm OB_B}} )\\
  \nonumber&& = P(({\bf R_{\rm OB_{Ai}}}\neq {\bf R_{\rm OB_{B1}}})~\&~({\bf R_{\rm OB_{Ai}}}\neq{\bf R_{\rm OB_{B2}}}),\\
  \nonumber&& ...,~\&~({\bf R_{\rm OB_{Ai}}}\neq {\bf R_{\rm OB_{Bp}}}) )\\
  \nonumber&&=P({\bf R_{\rm OB_{Ai}}}\neq {\bf R_{\rm OB_{\rm B1}}})P({\bf R_{\rm OB_{Ai}}}\neq {\bf R_{\rm OB_{\rm B2}}})\\
  &&...P({\bf R_{\rm OB_{Ai}}}\neq {\bf R_{\rm OB_{\rm Bp}}} )
  ,~i\in \{1,2, ...,p\}.
  \end{eqnarray}
  
  \begin{eqnarray}\label{eq:10}
  && P( {\bf R_{\rm OB_{Ai}}} \neq  {\bf R_{\rm OB_{Bj}}} )\\
  \nonumber&& = 1-{\it P}({\bf R}_{\rm OB_{Ai}} = {\bf R}_{\rm OB_{Bj}}) ~~\text{for}~i,j\in \{1,2, ...,p\}.
  \end{eqnarray}
  As for OB-PUFs, the obfuscated response consists of multiple bits, therefore, $ {\bf R_{\rm OB_{Ai}}} \neq  {\bf R_{\rm OB_{Bj}}} $ actually implies that the $ f_{\rm HD}(\bf R_{\rm OB_{Ai}},\bf R_{\rm OB_{Bj}})> {\it n}_{\rm mismatch} $. Consequently,
  \begin{eqnarray}
  \nonumber&& P( {\bf R_{\rm OB_{Ai}}} \neq  {\bf R_{\rm OB_{Bj}}} )\\
  \nonumber&&= 1-{\it P}(f_{\rm HD}(\bf R_{\rm OB_{Ai}},\bf R_{\rm OB_{Bj}})\le {\it n}_{\rm mismatch})\\
  &&=1-\sum\limits_{0}^{n_{\rm mismatch}}\frac{1}{2^{n_{\rm ins}}}.
  \end{eqnarray}
  
  Therefore,
  \begin{eqnarray}
  \nonumber&& P( {\bf R_{\rm OB_{Ai}}} \neq  {\bf R_{\rm OB_B}} )\\
  &&= (1-\sum\limits_{0}^{n_{\rm mismatch}}\frac{1}{2^{n_{\rm ins}}})^p~,~i\in \{1,2, ...,p\},
  \end{eqnarray}
  and,
  \begin{eqnarray}
  P( {\bf R_{\rm OB_A}} \neq  {\bf R_{\rm OB_B}} )\nonumber&&= ((1-\sum\limits_{0}^{n_{\rm mismatch}}\frac{1}{2^{n_{\rm ins}}})^p)^p\\
  &&=(1-\sum\limits_{0}^{n_{\rm mismatch}}\frac{1}{2^{n_{\rm ins}}})^{p^2}.
  \end{eqnarray}
  
  As $ {\hat p}_{\rm inter} = P( {\bf R_{\rm OB_A}} \neq  {\bf R_{\rm OB_B}} ) $, then $ {\hat p}_{\rm inter} $ of OB-PUFs can be formally expressed as:
  
  \begin{equation}\label{eq:Pr_dist}
  { {\hat p}_{\rm inter}}=(1-\sum\limits_{0}^{n_{\rm mismatch}}\frac{1}{2^{n_{\rm ins}}})^{p^2}.
  \end{equation}
  
  Based on (\ref{eq:Pr_dist}), we can see that smaller number of PUF instances $ n_{\rm ins} $ in a PUF block decreases $ {\hat p}_{\rm inter} $. For example, if $ n_{\rm ins} $ is one, then the $ {\hat p}_{\rm inter} $ is only $ \frac{1}{2^4} $ given $ p=2 $. This extremely small $ {\hat p}_{\rm inter} $ implies that the OB-PUF based authentication mechanism yields undesirably high FAR. Fortunately, inter-distance can be easily regained by increasing the number of PUF instances implemented in parallel, and simultaneously queried by the same full length challenge.
  \subsection{The Binomial Probability Estimator $ {\hat p}_{\rm intra} $ of OB-PUF}\label{Sec:BER}
  In general, the binomial probability estimator $ {\hat p}_{\rm intra} $ of a conventional PUF is actually the probability that two repeatedly generated 1-bit responses, given the same challenge applied to the same PUF, are different.
  
  As for an OB-PUF, the ${\hat p}_{\rm intra} $ is the probability that two repeatedly generated obfuscated responses given the same full length challenge applied to the same OB-PUF are different. Considering that an OB-PUF produces an $ n_{\rm ins} $-bit obfuscated response, we can see that $ {\hat p}_{\rm intra} $ is actually the probability of $ f_{\rm HD}({\bf R}_{\rm OB}^1,{\bf R}_{\rm OB}^2)>n_{\rm mismatch} $, where the $ {\bf R}_{\rm OB}^1 $ and $ {\bf R}_{\rm OB}^2 $ are two repeatedly generated obfuscated responses at two different time instances. However, it is much easier to first formulate the probability of $ f_{\rm HD}({\bf R}_{\rm OB}^1,{\bf R}_{\rm OB}^2)\le n_{\rm mismatch} $ given as:
  \begin{equation}\label{eq:21}
  \begin{split}
  \sum\limits_{i=0}^{n_{\rm mistmatch}}\binom{n_{\rm ins}}{i}\times (1-{{\hat p}_{\rm intra}^{\rm puf}})^{(n_{\rm ins}-i)}\times{({\hat p}_{\rm intra}^{\rm puf})}^i.
  \end{split}
  \end{equation}
  In (\ref{eq:21}), to distinguish the $ {\hat p}_{\rm intra} $ of OB-PUFs that of the underlying PUF in the PUF block, we refer to the $ {\hat p}_{\rm intra} $ of an underlying PUF as $ {\hat p}_{\rm intra}^{\rm puf} $. 
  
  Then, formally, we can express the binomial probability estimator of intra-distance of OB-PUF, $ {\hat p}_{\rm intra}$, as:
  \begin{eqnarray}\label{eq:FRR1RSimMat}
  &&{\hat p}_{\rm intra}\\
  \nonumber&&=1-P(f_{\rm HD}({\bf R}_{\rm OB}^1,{\bf R}_{\rm OB}^2)\le n_{\rm mismatch})\\
  \nonumber&&=1- \Bigg[\sum\limits_{i=0}^{n_{\rm mistmatch}}\binom{n_{\rm ins}}{i} (1-{{\hat p}_{\rm intra}^{\rm puf}})^{(n_{\rm ins}-i)}{({\hat p}_{\rm inter}^{\rm puf})}^i\Bigg].
  \end{eqnarray}
  \subsection{Evaluating OB-PUF Authentication Capability}\label{AuthenPower}
  \begin{table*} 
  \centering 
  \caption{The relationship between authentication power and configurations of OB-PUF($ n_{\rm ins} $, $ p $, $ n_{\rm mismatch} $).}
  \resizebox{0.8\textwidth}{!}{
  \begin{tabular}{c|| c c c c || c c c c || c c c c} 
  \toprule 
  \toprule 
  &\multicolumn{4}{c}{EER $ < 10^{-6} $} & \multicolumn{4}{c}{EER $ < 10^{-9} $}& \multicolumn{4}{c}{EER $ < 10^{-12} $} \\
  \cmidrule(l){2-5} \cmidrule(l){6-9} \cmidrule(l){10-13}
  OB-PUF($ n_{\rm ins} $, $ p $, $ n_{\rm mismatch} $) & $ n $ & $ {n}_{\rm EER} $ & FAR$ ^* $ & FRR$ ^* $ &$ n $ & $ {n}_{\rm EER} $ & FAR$ ^* $ & FRR$ ^* $ & $ n $ & $ {n}_{\rm EER} $ & FAR$ ^* $ & FRR$ ^* $  \\ 
  \midrule 
  OB-PUF(2, 2, 0)& 294 & 57 & $ -6.06 $ & $ -6.06 $ & 465 & 90 & $ -9.02 $ & $ -9.06 $ & 641 & 124 & $ -12.02 $ & $ -12.13 $ \\ 
  OB-PUF(4, 2, 0) & 219 & 46 & $ -6.05 $ & $ -6.01 $ & 348 & 73 & $ -9.01 $ & $ -9.05 $ & 478 & 100 & $ -12.03 $ & $ -12.00 $ \\ 
  OB-PUF(4, 4, 0)& 599 & 159 & $ -6.06 $ & $ -6.02 $ & 950 & 252 & $ -9.04 $ & $ -9.01 $ & 1308 & 347 & $ -12.03 $ & $ -12.06 $ \\ 
  OB-PUF(8, 4, 0)& 42 & 30 & $ -6.14 $ & $ -6.19 $ & 68 & 48 & $ -9.49 $ & $ -9.25 $ & 92 & 65 & $ -12.15 $ & $ -12.27 $ \\ 
  OB-PUF(8, 4, 1) & 58 & 15 & $ -6.22 $ & $ -6.19 $ & 90 & 23 & $ -9.14 $ & $ -9.02 $ & 125 & 32 & $ -12.15 $ & $ -12.27 $ \\ 
  OB-PUF(16, 4, 0) & 39 & 36 & $ -7.71 $ & $ -6.13 $ & 57 & 53 & $ -9.17 $ & $ -9.14 $ & 79 & 73 & $ -12.64 $ & $ -12.04 $ \\ 
  OB-PUF(16, 4, 1) & 15 & 12 & $ -6.43 $ & $ -6.27 $ & 24 & 19 & $ -9.22 $ & $ -9.54 $ & 32 & 25 & $ -12.11 $ & $ -12.16 $ \\ 
  \bottomrule 
  \end{tabular}}
  \begin{tablenotes}  
  \item[a] Note: the $ ^* $ symbol means value is from log10.
  
  \end{tablenotes}  
  \label{tab:AuthenTab} 
  \end{table*}
  We can quantitatively analyze the FAR and FRR of OB-PUF when using $ n $ OB-CRPs for authentication. According to (\ref{eq:FAR}) and (\ref{eq:FRR}), the FAR and FRR are determined by binomial probability estimators $ {\hat p}_{\rm inter} $ and $ {\hat p}_{\rm intra} $, chosen $ n_{\rm th} $ and the number of OB-CRPs $ n $. Notably, the binomial probability estimators of OB-PUFs, $ {\hat p}_{\rm inter} $ and ${\hat p}_{\rm intra} $, are functions of OB-PUF implementation parameters $ n_{\rm ins} $, $ p $ and $ n_{\rm mismatch} $. Hence, various combinations of these implementation parameters naturally lead to a family of OB-PUFs with differing physical features and security features such as authentication capability. Here, it is convenient to express the resulting family of OB-PUFs as OB-PUF($ n_{\rm ins} $, $ p $, $ n_{\rm mismatch} $).
  
  In Table~\ref{tab:AuthenTab}, we give a quantitative evaluation of authentication power under different configurations of OB-PUF($ n_{\rm ins} $, $ p $, $ n_{\rm mismatch} $). Here, $ {\hat p}_{\rm intra}^{\rm puf} $ is selected as 5\% since that is the worst-case $ {\hat p}_{\rm intra}^{\rm puf} $ reported for APUF exposed to $ 45\celsius $ variation of temperature and 2\% supply voltage variation according to the experimental data in~\cite{lim2005extracting}.
  \subsubsection{\bf Size of the PUF Block}We can see from Table~\ref{tab:AuthenTab} the required number of OB-CRPs, $ n $, to achieve a specific EER, e.g., $ 10^{-6} $, decreases as $ n_{\rm ins} $ increases. A larger number of PUF instances $ n_{\rm ins} $ requires a smaller $ n $, which in turn requires less time for sequentially sending and receiving those OB-CRPs and thus enabling faster authentication. However, this approach will result in higher overhead costs as the silicon area needed to implement an OB-PUF increases. In practical applications where authentication period is not of strict concern, selecting a small value for $ n_{\rm ins} $ provides a lightweight primitive.
  
  \subsubsection{\bf Number of Pattern Vectors}Based on the OB-PUF(4, 2, 0) and OB-PUF(4, 4, 0), we can see that as the number of pattern vectors $ p $ increases, the number of needed OB-CRPs $ n $ increases. This is because the larger $ p $ decreases the inter-distance according to (\ref{eq:Pr_dist}), but from a security perspective, a larger $ p $ achieves a higher security level, because it reduces the probability of an adversary correctly guessing the selected pattern vector, see Section~\ref{Analysis}. 
  \begin{figure}
  \centering
  \includegraphics[trim=0 0 0 0,clip,width=0.35\textwidth]{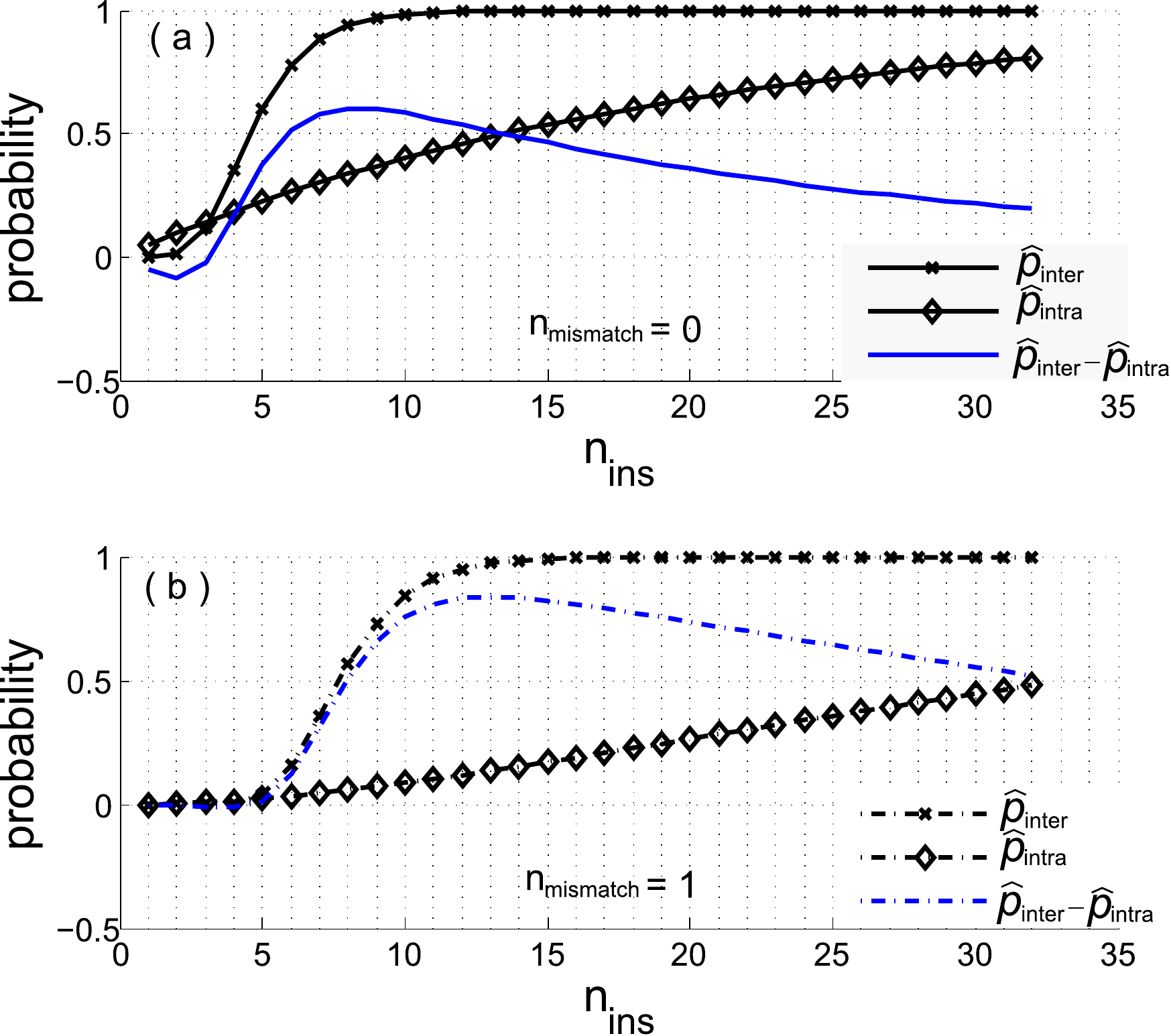}
  \caption{$ {\hat p}_{\rm inter} $ and $ {\hat p}_{\rm intra} $ as a function of $ n_{\rm ins} $ when (a) $ n_{\rm mismatch}=0 $ under a fixed $ p=4 $, and (b) $ n_{\rm mismatch}=1 $ under a fixed $ p=4 $.
  }
  \label{fig:FRRAndTAR_1R}
  \end{figure}
  
  \subsubsection{\bf Mismatch Tolerance}Considering OB-PUF(16, 4, 0) and OB-PUF(16, 4, 1), we can see that $ n$ can be smaller by using a slightly larger $ n_{\rm mismatch} $, e.g., $ n_{\rm mismatch}=1 $, recall that $ n_{\rm mismatch}$ is the number of tolerated mismatching bits when comparing an emulated obfuscated response with the received obfuscated response---see Determining a Match in Section~\ref{Sec:AuthenSecheme}. In detail, setting $ n_{\rm mismatch}=1 $ increases the difference between $ {\hat p}_{\rm inter} $ and $ {\hat p}_{\rm intra} $ when $ n_{\rm ins} $ is large, for example, more than ten as illustrated in Fig.~\ref{fig:FRRAndTAR_1R}, and in turn expedites the authentication process. In other words, the authentication period can be made shorter. Notably, for $ n_{\rm ins}=16 $ for both OB-PUF(16, 4, 0) and OB-PUF(16, 4, 1), the authentication period benefits from setting $ n_{\rm mismatch}=1 $. This improvement can also be understood from Fig.~\ref{FARFRR}. Consider a fixed $ n $ and fixed $ n_{\rm th} $, when the difference between $ {\hat p}_{\rm inter} $ and $ {\hat p}_{\rm intra} $ is larger---in other words, difference between intra-distance and inter-distance is increased, FAR and FRR will eventually decrease.
  
  Overall, OB-PUFs can be flexibly configured to suit different application scenarios according to their performance requirements such as time to complete authentication, level of security and hardware cost constraints. If the authentication time is a concern, then using a large $ n_{\rm ins} $ and setting $ n_{\rm mismatch}=1 $ or even larger is possible. Otherwise, if the OB-PUF integrated device is resource constrained and the authentication time is not a strict concern, then using a small $ n_{\rm ins} $, e.g., two, and setting $ n_{\rm mismatch}=0 $ is possible.
\subsection{Reconfiguring Latent Pattern Vectors}\label{Sec:ReconfigurePattern}
\begin{figure}
\centering
\includegraphics[trim=0 0 0 0,clip,width=0.50\textwidth]{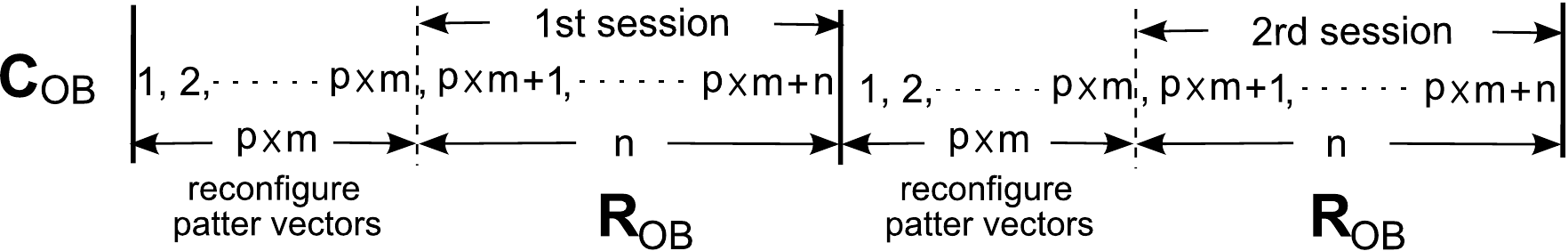}
\caption{Pattern vectors are reconfigured before authentication session starts.}
\label{fig:SessionDepict}
\end{figure}
Now we introduce how we reconfigure the pattern vectors during an authentication phase. We eschew reliance on NVMs to store reconfigured pattern vectors by generating them on demand from the underlying APUFs within the pattern vector reconfiguring block, as shown in Fig.~\ref{ChallengePreProcessing}. Hence, the pattern vectors are only known by the server and therefore hidden from other parties. 

This paper focuses on reconfiguring the inserted values of the pattern vectors, as shown in Fig.~\ref{InsertBitValues}, used to form the full length challenge. It is reminded that all other information of the pattern vector such as the inserted positions and the strings used to obfuscate the responses can also be reconfigured in a similar manner. As depicted in Fig.~\ref{fig:SessionDepict}, before each authentication session consisting of $ n $ OB-CPRs, the inserted values with bit-length of $ p\times m $ are first generated from the underlying APUFs within the pattern vector reconfiguring block. The pattern vector reconfiguring block implements logic to assess all sequentially applied  $ p\times m $ challenges are different. If any two challenges out of  $ p\times m $ challenges are different, then those bits are stored in internal registers to be used for the current session. Otherwise, random numbers are filled into the registers. We consider that the APUFs in charge of reconfiguring pattern vectors share the same $ {\bf C}_{\rm OB} $ with the APUFs underlying the PUF blocks to simplify the logic implementation in practice, the difference is that the former APUFs only have $ k-m $ stages. In other words, the $ {\bf C}_{\rm OB} $ is the full length challenge of the APUFs underlying the reconfiguring pattern vectors block.

\subsubsection{Preselection of Partial Challenges for Pattern Vector Generation.}

In general, the naturally noisy feature of PUF responses prevents the stable generation of the inserted values and on-chip error correction must be applied. However, when a basic APUF is modeled/learned through a method such as support vector machines (SVM) \cite{xu2016using}, the predicted time delay $ t_{\rm dif} $\footnote{The predicted value does not contain the time unit, but it is linear to the actually delay time as detailed in \cite{xu2016using}} in Eq(\ref{Eq:t_dif})---standing for the difference in delay time between the top path and the bottom path of an APUF---contains two useful information: i) sgn($ t_{\rm dif} $) that determines the binary response; ii) the reliability of this response. If the $ t_{\rm dif} $ is alway far from zero, then a response from such a challenge can be reproduced with full confidence. Recent work of Xu {\it et al.} \cite{xu2016using} shows that judiciously determined challenges based on a learned statistical APUF model can generate error free responses even across a wide range of operating conditions as well as aging effects. 

Therefore, the server is able to easily determine $ p\times m $ partial challenges $ {\bf C}_{\rm OB} $---noting that the $ {\bf C}_{\rm OB} $ is the full length challenge of the APUFs with $ k-m $ stages underlying the pattern vector reconfiguring block---that always produce error-free responses on demand. We adopt such an error free response selection method to refresh the inserted values for reconfiguring all pattern vectors at the beginning of each authentication session. It is also noticed that such a reliable response selection method can further be employed to preselect partial challenges that are used for the following $ n $ OB-CRPs to enable an improved reliability of the OB-CRPs.
  
\section{Security Evaluations}\label{Analysis}
\subsection{Adversary Model}
We employ a common adversary model described in~\cite{ruhrmair2013puf,ruhrmair2010modeling,rostami2014robust} when evaluating OB-PUFs' security. In summary, the adversary is allowed to eavesdrop on the communication channel and arbitrarily apply challenges using the publicly accessible OB-PUF interface to collect an arbitrary number of CRPs. The adaversary's goal is to learn an accurate model of the OB-PUF using modeling attacks. 

We focus on recently revealed CMA-ES based modeling attacks where a direct relationship between a challenge and a response is not required~\cite{becker2015pitfalls}.Further, CMA-ES attacks have succeeded where previous learning method, e.g., LR, has failed, for example, the Slender PUF~\cite{becker2015pitfalls}.

\subsection{Modeling Attack Test Setup}\label{Sec:AttSetup}
All the CMA-ES attacks are executed using MATLAB 2012b, where the CMS-ES code is adopted from \cite{hansen2016cma}, and the processor has an Intel i7-3770CPU@3.4GHz CPU and a 16GB memory.
\subsubsection{OB-CRP Generation}
Similar to studies in ~\cite{ruhrmair2013puf,ruhrmair2010modeling,majzoobi2008testing,becker2015gap,becker2015pitfalls}, the CRPs of underlying APUFs are generated through simulations that can effectively model a physical APUF architecture. The delay values for each stage in an APUF are randomly produced following a standard normal distribution. When a specific response is required, a given challenge $ \bf C $ is applied to the APUF to determine which delay values $ {\sigma}_i^{0/1} $ in each stage will be selected according to the logic value of $ c_i $. Then, according to the linear additive delay model, the selected delay values of two electrical signal paths are added up, separately, and compared with each other to generate a response bit. Artificially injected noise is not used in the following attack tests. Therefore, a purely `logic' security of PUF design is evaluated. In this way, the evaluated security represents an upper bound on OB-PUF's resilience to modeling attacks~\cite{ruhrmair2013puf}.

As for generation of an OB-CRP for OB-PUFs, first, a partial challenge $ {\bf C}_{\rm OB} $ is randomly generated. Next, one of $ p $ possible full length challenges $ \bf C $ is formed according to a challenge obfuscation vector $ {\bf S}_C $---determined by a specific randomly chosen pattern vector as illustrated in Fig.~\ref{InsertBitValues}. Then its corresponding response $ \bf R $ is produced by applying the formed full length challenge to the PUF block. As illustrated in Fig.~\ref{InsertBitValues}, the obfuscated response $ {\bf R}_{\rm OB} $ is obtained by XORing a specific response obfuscation vector $ {\bf S}_R $ with $ \bf R $.

\subsubsection{Criterion for Breaking OB-PUFs}
In general, to successfully break a PUF, the prediction error rate of the learnt PUF model should be less than the $ {\hat p}_{\rm intra} $ of the PUF. In other words, the prediction accuracy of the learnt model should be higher than $ 1-{\hat p}_{\rm intra} $. We refer to this minimum prediction accuracy as $ P_{\rm min} $ and define it formally for OB-PUF below.
\begin{definition}{\bf Minimum prediction accuracy requirement.} \label{def:Ptarget}
The  $  P_{\rm min} $ is the minimum prediction accuracy of the learned PUF model that can be considered to successful break a PUF.
\end{definition}
For OB-PUFs, based on (\ref{eq:FRR1RSimMat}), $ P_{\rm min}=1-{\hat p}_{\rm intra} $, then $ P_{\rm min} $ is expressed as:
\begin{equation}\label{eq:Ptarget}
\begin{split}
{P_{\rm min}}=\sum\limits_{i=0}^{n_{\rm mistmatch}}\binom{n_{\rm ins}}{i}\times (1-{{\hat p}_{\rm intra}^{\rm puf}})^{(n_{\rm ins}-i)}\\\times{({\hat p}_{\rm intra}^{\rm puf})}^i,
\end{split}
\end{equation}
where $ {\hat p}_{\rm intra}^{\rm puf} $ is the binomial probability estimator of intra-distance distribution of the underlying PUF, e.g., APUF, in the PUF block. In the following studies in Section~\ref{Analysis}, we use the worst-case $ {\hat p}_{\rm intra}^{\rm puf}=5\% $ obtained when an APUF is exposed to 45$ \celsius $ variation in temperature and 2\% variation in supply voltage~\cite{lim2005extracting}.

Then the aim of an adversary is to build a model of an OB-PUF with predication accuracy that is higher than the $ {P_{\rm min}} $. If this is achieved, the OB-PUF is broken.

\subsection{CMA-ES Attacks on Reconfigurable OB-PUF}
Though there are variants of ES algorithms, we follow Becker's work~\cite{becker2015gap,becker2015pitfalls} using covariance matrix adaptation ES (CMA-ES)~ \cite{hansen2016cma}. 

Overall, there are three fitness metrics in the literature~\cite{becker2015gap,becker2015pitfalls} to achieve ES attacks, they are based on: i) response Hamming distance, ii) response Hamming weight and iii) response reliability. Choosing a fitness function or metric is a key challenge in mounting ES attacks. We start from the response Hamming distance based fitness metric following Becker's work in~\cite{becker2015gap}.
\subsubsection{Response Hamming Distance based Fitness Metric}
To ease understanding of evaluating the fitness of a candidate delay time vector, ${\boldsymbol \omega} $, when the reconfigurable pattern vectors are employed, we first describe how to define a fitness function under the assumption that pattern vectors are fixed and publicly known. For each OB-CRP, by knowing all pattern vectors, an adversary forms all $ p $ possible full length challenges, $ {\bf C}_{i1},..., {\bf C}_{ip} $, and, all possible obfuscated responses $ {\bf R}_{\rm OB_{i1}}^{\prime},..., {\bf R}_{\rm OB_{ip}}^{\prime} $ are computed according to the selected ${\boldsymbol \omega} $ and using all the known pattern vectors. Now, the minimum FHD between all possible computed obfuscated responses and the under-attack OB-PUF's obfuscated response is evaluated:
\begin{equation}\label{Eq:FittestFun}
f_i=\min_{j=1,...,p} \{f_{\rm HD}({\bf R}_{\rm OB_{ij}}^{\prime},{\bf R}_{\rm OB_i}\}).
\end{equation}
Then the fitness metric $ f $ to evaluate the goodness of candidate, ${\boldsymbol \omega} $, is acquired by summing all $ f_i $ of all OB-CRPs; smaller the $ f $, fitter the learned PUF model~\cite{becker2015pitfalls}.

Notably, the ES algorithm needs to iterate many generations before reaching the best delay time vectors $ {\boldsymbol \omega} $. In each generation, many children are produced; for each child, it has a corresponding delay time vector inherited from its parent(s). 

Now considering the implemented reconfigurable pattern vectors during authentication phase, in order to evaluate the fitness function in Eq (\ref{Eq:FittestFun}), for each child in each generation, and for each authentication session, the ES algorithm has to first work out the pattern vector by predicting, then form all possible full length challenges, all based on a delay time vector of each child in every generation. At the end, all $ f $ for all sessions are summed up and normalized to form a fitness metric that is applicable to the reconfigurable OB-PUF.

The crux of reconfigurable pattern vector is the stringent requirement of predicting correct inserted values by an adversary. The full length challenge feature has a very high chance of being greatly different from the correct one if any bit of the inserted values is erroneous. This can be observed from Eq (\ref{Eq:ChallengeFeature}), where the challenge feature is a function of all bits in the challenge vector. It seems that arbitrarily measuring the OB-PUF by the adversary may not be a good choice to collect OB-CRPs even when the physical measurements on OB-PUF is allowed. Because, the adversary is unable to guarantee the $ p\times m $ inserted values being error freely generated when arbitrary partial challenges $ {\bf C}_{\rm OB} $ are applied. The $ p\times m $ error free inserted values are foundations for correctly evaluating the rest $ n $ OB-CRPs. If the inserted values, thus pattern vectors, are incorrect, then during the ES attack, the incorrect OB-CRPs lead to misleading fitness assessments of candidate OB-PUF models. 

In addition, the reconfiguring pattern vector makes the attack computation time significantly to be increased. With regarding to a fixed public known pattern vector as in our conference design~\cite{gao2016obfuscated}, the ES algorithm is only required to compute all possible full length challenges and then form their corresponding challenge features once. The same challenge features are applicable for each child in each generation. When the reconfigured pattern vector is implemented, such a one-time challenge features computation is no longer applicable. For each session, each child in each generation, the challenge features vary. Therefore, ES algorithm has to recompute the refreshed pattern vector first, then form all possible full length challenge per authentication session, then computes their corresponding challenge features. Such a re-computation of reconfigured pattern vectors and then challenge features per session is unavoidable for any child in any generation of the ES algorithm. This recomputation results  into a significantly increased computation time of the ES attacks when the reconfiguring pattern vector per session is applied.

In the following, we evaluate the security of reconfigurable OB-PUFs under CMA-ES attacks.

{\bf OB-PUF(2,2,0).}
\begin{table}
\centering 
\caption{Prediction Rate of OB-PUF(2, 2, 0)}
\resizebox{0.40\textwidth}{!}{
\begin{tabular}{c| c | c | c | c | c | c} 
\toprule 
$ \# $row&\begin{tabular}{@{}c@{}} XORs\end{tabular} & \begin{tabular}{@{}c@{}} $ N_{\rm session} $ \end{tabular}& \begin{tabular}{@{}c@{}}$ N_{\rm generation} $ \end{tabular}& \begin{tabular}{@{}c@{}} $ n $ \end{tabular}&\begin{tabular}{@{}c@{}} $ m $ \end{tabular}&\begin{tabular}{@{}c@{}} $ P_{\rm pred} $; Time \end{tabular}\\ 
\midrule 
$ \# $1 & 2 & 100 & 100 & 300 & 3& 77.02\%; 05:23:11\\ 
$ \# $2 & 2 & 50 & 100 & 300 & 8& 68.54\%; 02:36:02\\
$ \# $3 & 2 & 100 & 100 & 300 & 8& 68.81\%; 05:14:32\\
$ \# $4 & 2 & 100 & 100 & 600 & 8& 68.71\%; 10:39:06\\
$ \# $5 & 2 & 100 & 400 & 300 & 6&
74.52\%; 20:54:03 \\
\bottomrule 
\end{tabular}}
	\begin{tablenotes}  
		\item[a] NOTE: i) XORs means the number of APUFs in a XOR-APUF within the pattern vector reconfiguring block to produce pattern vectors. ii) Time format is hour:minute:second. iii) The number of child in each generation uses the default setting of CMA-ES algorithm~\cite{hansen2016cma}.
	\end{tablenotes}
\label{tab:ESOB-PUF2P2} 
\end{table}
We first test OB-PUF(2,2,0), where two APUFs responses are XORed within the pattern vector reconfiguring block to reconfigure pattern vectors per session. Attack results are shown in Table \ref{tab:ESOB-PUF2P2}. Considering the $ P_{\rm min}=90.25\% $ and the $ P_{\rm pred} $ is around 77\%, the OB-PUF(2,2,0), hence, has not yet been broken. Recall that the $ m $ is the eliminated bits in the full length challenge. The results indicate that a larger $ m $ leads to a lower $ P_{\rm pred} $. This relies on the fact that the adversary's model has a lower probability of correctly predicting all $ p \times m $ inserted values all correct when the $ m $ increases. Prediction errors of the computed $ p \times m $ inserted values change the full length challenge feature and consequently increase erroneousness in the final computed obfuscated responses. This explains the decreased $ P_{\rm pred} $ from 77.02\% when $ m=3 $ to 68.81\% when $ m=8 $. When a larger  $ m $ such as 6 is used, the predict accuracy decreases greatly. We can observe this based on $ \# $row 1 and $ \# $row 5. From the ES learning perspective, the reason lies in the hardness of correctly determining all $ p \times m $ inserted values, where an inaccurate determination ~\cite{becker2015pitfalls}.

In addition, under expectations, the computation time is significantly increased compared with attacking fixed pattern vectors that only costs around 15 minutes. When $ N_{\rm session}=100 $, $ N_{\rm generation}=100 $ and $ n=300 $---this is the default setting when performing authentication as quantified in Table~\ref{tab:AuthenTab}, the computation time is significantly prolonged to more than five hours as shown in $ \# $row3.

{\bf OB-PUF(4,4,0).}
\begin{table}
\centering 
\caption{Prediction Rate of OB-PUF(4, 4, 0)}
\resizebox{0.40\textwidth}{!}{
\begin{tabular}{c| c | c | c | c | c | c} 
\toprule 
$ \# $row &\begin{tabular}{@{}c@{}} XORs\end{tabular} & \begin{tabular}{@{}c@{}} $ N_{\rm session} $ \end{tabular}& \begin{tabular}{@{}c@{}}$ N_{\rm generation} $ \end{tabular}& \begin{tabular}{@{}c@{}} $ n $ \end{tabular}&\begin{tabular}{@{}c@{}} $ m $ \end{tabular}&\begin{tabular}{@{}c@{}} $ P_{\rm pred} $; Time \end{tabular}\\ 
\midrule 
$ \# $1 & 1 & 100 & 100 & 600 & 3 & 76.42\%; 14:03:36\\ 
$ \# $2 & 2 & 50 & 100 & 600 & 3 & 55.74\%; 08:30:26\\ 
$ \# $3 & 2 & 50 & 100 & 600 & 6 & 52.00\%; 07:59:05\\ 
$ \# $4 & 2 & 100 & 100 & 600 & 3 & 56.87\%; 16:48:22\\ 
$ \# $5 & 2 & 100 & 100 & 600 & 6 & 54.42\%; 15:34:58\\
$ \# $6 & 2 & 100 & 200 & 600 & 3 & 60.32\%; 33:12:38\\ 
$ \# $7 & 2 & 100 & 200 & 600 & 6 & 55.90\%; 32:43:22\\
$ \# $8 &2 & 100 & 300 & 600 & 3 & 62.40\%; 50:10:56\\ 
$ \# $9 & 2 & 100 & 300 & 600 & 6 & 58.10\%; 49:46:33\\ 
\bottomrule 
\end{tabular}}
\label{tab:ESOB-PUF4P4} 
\end{table}
We now test OB-PUF(4,4,0). 
The test starts with using only a single APUF to produce reconfigured pattern vectors. In this context, the $ P_{\rm pred} $ is up to $ 76.42\% $ as shown in $ \# $row1 that is close but still lower than the $ P_{\rm min} $ of 81.45\% according to Eq(\ref{eq:Ptarget}). When a XOR2-APUF within the pattern vector reconfiguring block is used, the $ P_{\rm pred} $ is significantly reduced to $ 56.87\% $, see $ \# $row4. The reduced $ P_{\rm pred} $ is mainly attributed to the increased erroneousness that occurs during the determination of pattern vectors by the adversary's model as a result of the XORing operation. On the other hand, this result agrees well with Becker's view \cite{becker2015pitfalls} that an obfuscated PUF construction such as the Slender PUF or the OB-PUF in conjunction with PUFs already resistant to traditional ML attacks increases its resistance to CMA-ES attacks. The side product of increased computation time overhead is confirmed again by observing the last column.

\begin{figure}
\centering
\includegraphics[trim=0 0 0 0,clip,width=0.45\textwidth]{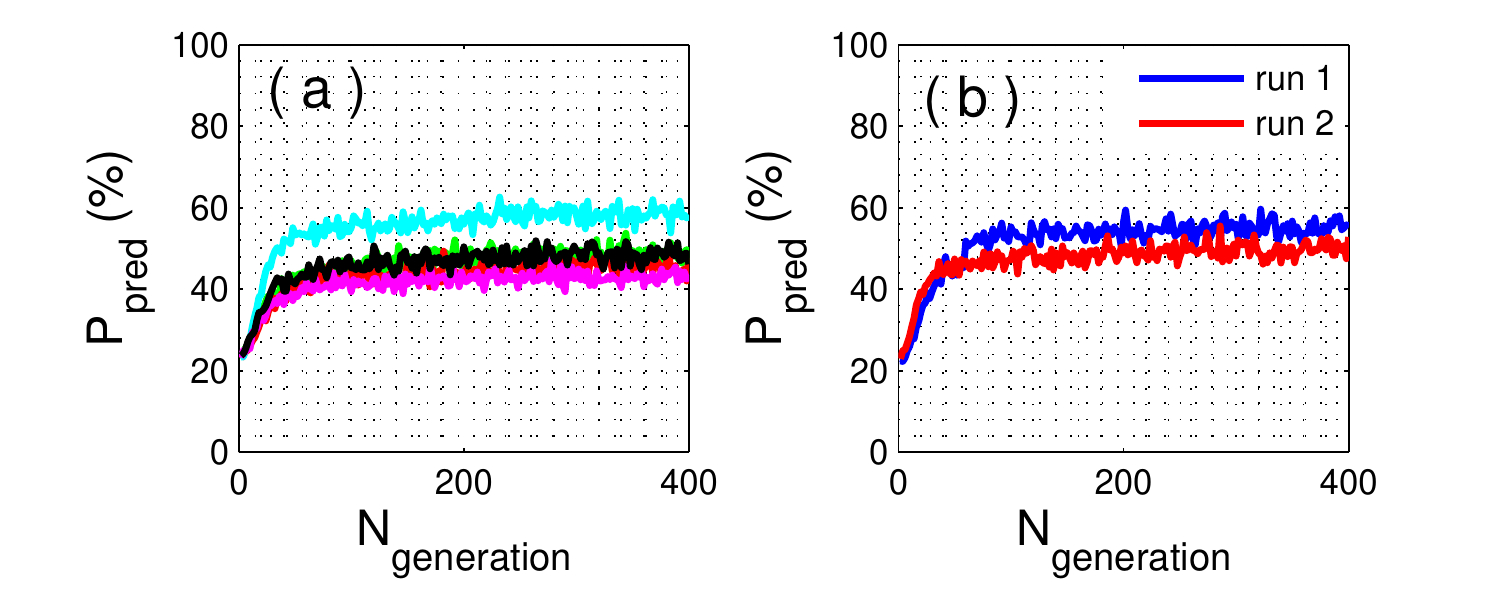}
\caption{The y-axis shows the achieved prediction accuracy $ P_{\rm pred} $ of the CMA-ES attacks. The targeted accuracy $ P_{\rm min} $ is around 81.45\% to break the OB-PUF(4,4,0). (a) Progression of five runs of the CMA-ES attacks. For all runs, $ N_{\rm session}=100 $, $ n=600 $ and $ m=6 $ are used. Each run takes around 67 hours. (b) For run 1,  $ N_{\rm session}=100 $, $ n=1200 $ and $ m=6 $ are used. For run 2,  $ N_{\rm session}=200 $, $ n=600 $ and $ m=6 $ are used. Each run takes around 140 hours.}
\label{fig:P_predPlot}
\end{figure}

In Fig.~\ref{fig:P_predPlot} (a), it depicts the progression of fives runs of the CMA-ES on the OB-PUF(4,4,0). For all runs, $ N_{\rm session}=100 $, $ n=600 $ and $ m=6 $. We can see, for all runs, the $ P_{\rm pred} $ is converges within 100 generations with the first 50 generations leading to an increased $ P_{\rm pred} $. After that, increasing the generations does not help improve the $ P_{\rm pred} $. This lies on the $ p\times m $ inserted values, reconfigured per session, are hard to be computed all correctly by an adversary, which eventually misdirects the ES evaluation and disrupts further optimization~\cite{becker2015pitfalls}. Each run is up to 400 generations and takes around 67 hours. The $ P_{\rm pred} $ is still far below the targeted $ P_{\rm min}=81.45\% $, which validates the significantly resilience to CMA-ES attacks when the pattern vector reconfiguration is implemented. Fig.~\ref{fig:P_predPlot} (b) investigates the $ P_{\rm pred} $ when the $ N_{\rm session} $ or $ n $ is increased, more specifically, more OB-CRPs are exploited for training. Using more OB-CRPs has no, at least negligible, improvement of adversary's model predication accuracy, which confirms one more time that reconfigurable latent pattern vectors misdirect the ES optimization.

We have extensively evaluated the OB-PUF's security under the ES attacks using the response Hamming distance based fitness metric. Next we analyze the response Hamming weight and response reliability based fitness metrics, respectively.

\subsubsection{Response Hamming Weight based Fitness Metric} CMA-ES attacks using response Hamming weight fitness function have broken the security of the slender PUF~\cite{becker2015pitfalls}. Though OB-PUF and Slender PUF both obfuscate the direct relationship between a challenge and its response, there is a major difference between the OB-PUF and Slender PUF that eliminates the OB-PUF's vulnerability to CMA-ES attacks employing the Hamming weight based fitness test. In the Slender PUF protocol, each authentication session shares the same random indice that is, to be simplified, alike the random indice determining which pattern vector is chosen in the OB-PUF. But in OB-PUF, the random indice changes per OB-CRP rather than per session as in the Slender PUF protocol. In addition, to properly evaluate the Hamming weight based objective function, one premise is that the $ n_{\rm ins} $ should be large enough in order to greatly outperform the noise effects (equivalent to a high signal to noise ratio, in short, SNR), which has been pointed out and also verified in \cite{becker2015pitfalls}. Generally, to help evaluate the Hamming weight fitness performance, a $ n_{\rm ins} $ that is in hundreds magnitude is necessary, however, the very small $ n_{\rm ins} $ of OB-PUF is far way from performing the CMA-ES attacks based on the Hamming weight fitness that is applicable to the Slender PUF. Please also note that the Hamming weight of the obfuscated response $ {\bf R}_{\rm OB} $ is already elaborately designed to be balanced, detailed in Section~\ref{Sec:ResponseOB}, this further eliminates the concern of Hamming weight fitness test enabled CMA-ES attacks. 

\subsubsection{Response Reliability based Fitness Metric}\label{Sec:HWESAttack} In~\cite{becker2015gap}, the reliability-based CMA-ES attack is applicable to XOR-APUFs. A proper reliability based fitness to mount CMA-ES attack on the OB-PUFs has not been figured out. If somehow, a proper reliability based fitness can be found for OB-PUF. This attack is still difficult to the reconfigurable OB-PUF. The main reason is that the correct determination of inserted values and then full length challenge feature is still required before performing such a reliability fitness based ES attacks. In other words, even the response of `1'/`0' is unnecessary, which full length challenge presents an unreliable response is still required~\cite{becker2015pitfalls}.

\subsection{Discussion} 
It was recognized that employing APUFs as building blocks becomes very challenging in front of CMA-ES attacks~\cite{becker2015pitfalls} even that APUFs has extremely attractive properties such as compact structure and, most importantly, large CRP space. The reconfigurable OB-PUF serves an initial investigation of preventing powerful ES attacks. From the designing of reconfigurable OB-PUF, three hints are learned for future more efficient constructions: i) forcing the ES to compute such as the challenge feature in each child and in each generation, which exploits the uncircumventable process of the ES algorithm to significantly increase its computation time, ideally to be exponentially increased; ii) increasing the number of APUFs in a XOR-APUF, e.g., more than two, within the pattern vector reconfiguring block can significantly increase the OB-PUF's ML-resilience further, this has been validated in Table~\ref{tab:ESOB-PUF4P4}. Or increasing the number of $ p\times m $ such as such as $ p=4,~m=18 $ to exponentially decrease the capability of the adversary's model predicting all pattern vectors all correct as validated in Table~\ref{tab:ESOB-PUF2P2} and Table~\ref{tab:ESOB-PUF4P4}; iii) fully exploiting the uneven access to the underlying PUFs would be a plausible tool to exploit asymmetrical information obtained between the trusted party, e.g., the server and other party, e.g, adversary. For example, the server has the capability of generating error free responses on demand based on the learned statistical model of a basic APUF~\cite{xu2016using}.

\section{Conclusion}\label{Conclusion}
In this paper, we proposed a reconfigurable latent obfuscation technique that results into the design of reconfigurable OB-PUFs. This is a continuous investigation of a strong PUF scheme that enables the most desirable PUF-oriented lightweight entity authentication either using error correction logic nor a cryptographic algorithm, which has been keeping pursued for over a decade in the PUF community~\cite{yulockdown,delvaux2015survey}.

The OB-PUF prevents an adversary carrying out successful modeling attacks due to that both pattern vectors and their selection to form possible full length challenges are unknown, but still allows a server to successfully authenticate an OB-PUF. In addition, pattern vectors are also reconfigured on demand and act as one-time pads per authentication session. We implement the most powerful modeling attacks to date, CMA-ES attack, to evaluate reconfigurable OB-PUFs security through extensive case studies, and demonstrate the significant modeling attack resilience. The reconfigurable OB-PUF studied in this paper serves as an very initial PUF designs that show significantly increased resilience to ES attacks without constraints on the available authentication rounds as in previous work~\cite{yulockdown}. We also summarize several useful hints to suggest future PUF designers to propose more efficient ES attacks resilient PUF constructions.

\section{Acknowledgment}
This research was supported by a grant from the Aus-
tralian Research Council (DP140103448). The authors would
also like to thank the sponsorship from China Scholarship
Council (201306070017).

\end{document}